\documentclass[12pt,amsmath,amssymb,epsf,epic,cite]{article}

\usepackage{amsmath}
\usepackage{amssymb}
\usepackage{graphicx}
\usepackage{color}

\numberwithin{equation}{section}

\newcommand{\e}{\,{\rm e}}
\newcommand{\blue}[1]{{\color{black} #1}}

\newcounter{resultcounter}[section]

\newcommand{\scalprod}[2]{\left\langle {#1}, {#2}\right\rangle}
\newcommand{\bbbone}{\mathchoice {\rm 1\mskip-4mu l} {\rm 1\mskip-4mu l}{\rm 1\mskip-4.5mu l} {\rm 1\mskip-5mu l}}

\renewcommand{\r}{{\rm R}}
\newcommand{\s}{{\rm S}}

\renewcommand{\i}{{\rm i}}

\renewcommand{\i}{{\rm i}}
\renewcommand{\r}{{\rm R}}

\begin{document}

\title{ Quantum Markovian master equations:\\
resonance theory \blue{shows validity for all time scales}}
\author{Marco Merkli \footnote{merkli@mun.ca, https://www.math.mun.ca/$\sim$merkli/}\\	
Department of Mathematics and Statistics  \\
Memorial University of Newfoundland\\
St. John's, 
Canada A1C 5S7}

	\maketitle

\begin{abstract}
Quantum systems coupled to environments exhibit intricate dynamics. The master equation gives a Markov approximation of the dynamics, allowing for analytic and numerical treatments. It is ubiquitous in theoretical and applied quantum sciences.  The accuracy of the master equation approximation was so far proven  \blue{for small values of the system-environment interaction coupling strength $\lambda$, under the additional constraint that time $t$ must not exceed an upper bound, $\lambda^2 t\le$ constant.} Here, we show that the Markov approximation is valid for fixed \blue{small} coupling strength and for all times.  We also construct a new approximate Markovian dynamics -- a completely positive, trace preserving semigroup -- which is asymptotically in time exact, to all orders in the coupling.
\end{abstract}



\maketitle

\blue{
\section{Introduction}

The evolution in quantum theory is governed by the Schr\"odinger equation. When a system is coupled its environment, the Schr\"odinger equation applies to the whole system-environment complex. The effective evolution of degrees of freedom of the system, {\em i.e.}, the open system dynamics, does not follow a  Schr\"odinger equation, though. Finding this effective equations is difficult. Under suitable conditions, in particular for weak system-environment coupling and if the environment has correlations which decay sufficiently quickly in time, one expects this evolution to be approximately Markovian. The corresponding equation is the ubiquious Markovian master equation. In this paper, we show how to obtain bounds for the accuracy of this approximation in a rigorous way. We do this using the so-called dynamical resonance theory. Our method works under certain hypotheses specified below and, to our knowledge, it is the only one able to derive such rigorous bounds.   
\medskip

Advantages of our approach are:

\begin{itemize}
\item[$\bullet$] We give rigorous bounds for the accuracy of the Markovian approximation, valid for all times. In particular, we show that the usual master equation generated by the Davies generator, is accurate to $O(\lambda^2)$, independently of time $t$ for all $t\ge 0$, where $\lambda$ is the system environment coupling constant. So far, this accuracy was shown to hold only under the additional constraint $\lambda^2t \le {\rm constant}$. 

\item[$\bullet$] We construct another, new Markovian approximation which is asymptotically exact. That is, for which the final state is the correct reduced equilibrium state of the system, to all orders in the perturbation parameter $\lambda$. This is an improvement over the approximation based on the Davies generator, since the latter predicts a final state which deviates from the true one by $O(\lambda^2)$.

\item[$\bullet$]  Our method works for initial system-environment states which are entangled. The techniques developed in most of the literature, only works under the assumption of disentangled system-environment initial states. Our method also describes the evolution of observables of the environment, but we do not elaborate on this aspect in the current paper.

\end{itemize}

Difficulties we encounter in our approach are:

\begin{itemize}
\item[$\bullet$] The error bounds for the accuracy of the Markovian approximations we derive involve constants. Those constants do not depend on the system-environment coupling strength $\lambda$ nor on time $t$. However, they depend on other parameters, such as the dimension of the system and the smoothness and infrared and ultraviolet behaviour of the coupling function. We have not obtained the explicit dependences on these parameters (even though in principle, it is possible to do).

\item[$\bullet$] The environment, also called reservoir, consists of free quantum particles. In the present work, we consider bosons. It is possible to take a reservoir of free fermions, but a reservoir of interacting particles is not treatable, up to now.
\end{itemize}

In order to be able to focus on the main ideas of the dynamical resonance theory, we make the technically most advantageous assumptions in this paper. This means the class of systems we  treat here is somewhat restricted by stronger regularity assumptions. However, we will extend the theory in several directions, including the following:

\begin{itemize}
\item[$\bullet$]  The assumptions we make in this manuscript imply that reservoir correlations decay exponentially quickly in time. This is not necessary for the dynamical resonance theory to work. By using Mourre theory (as opposed to complex spectral deformation), we will treat the situation where correlations are merely polynomially decaying. 

\item[$\bullet$] We plan to discuss in detail the system evolution for initially entangled system-environment states. In the current manuscript, we set up the resonance theory for the general, possibly entangled  case (Section \ref{secrepdyn}), but we discuss finer detail on the dynamics, namely our Results 1-3 (Sections \ref{subs1}-\ref{subs3}) only for factorized initial states. 

\end{itemize}

}

\section{Main results}
\label{Sec2}

\blue{
\subsection{The model}
}

We consider open quantum system Hamiltonians
\begin{equation}
	H = H_\s +H_\r +\lambda G\otimes\varphi(g)
	\label{1}
\end{equation}
where $H_\s$ is an $N\times N$ hermitian matrix with eigenvalues $E_j$ and eigenvectors $\phi_j$,
\begin{equation}
	\label{2}
	H_\s = \sum_{j=1}^N E_j |\phi_j\rangle\langle\phi_j|
\end{equation}
and $H_\r$ is the environment, or reservoir Hamiltonian
\begin{eqnarray}
	\label{3}
	H_\r = \sum_{k} \omega_k a^*_ka_k,
\end{eqnarray}
describing modes of a collection of harmonic oscillators, labelled by $k$. Their frequencies are $\omega_k>0$ (we set $\hbar=1$) and the creation and annihilation operators $a^*_k$, $a_k$, satisfy the canonical commutation relations $[a_k,a^*_\ell] = \delta_{k,\ell}$ (Kronecker symbol). The interaction term contains a coupling constant $\lambda\in\mathbb R$, an interaction operator $G$ (hermitian $N\times N$ matrix), and it is linear in the field operator 
\begin{equation}
	\label{4}
	\varphi(g) = \frac{1}{\sqrt 2}\sum_{k} g_k a^*_k + {\rm h.c.},
\end{equation}
where ${\rm h.c.}$ denotes the hermitian conjugate.  The collection of the numbers $g_k\in \mathbb C$ constitutes the form factor $g$. The size of $g_k$  determines how strongly the mode $k$ is coupled to the system. 

To describe irreversible effects \blue{-- such as thermalization and decoherence in the small system --} it is necessary to pass to a limit where the oscillator frequencies $\omega_k$ take on {\em continuous} values (and hence so must $k$). In principle, the parameter $k$ belongs to an arbitrary continuous set. For instance, having in mind a reservoir modeling a (scalar) quantized field in physical space ${\mathbb R}^3$ (infinite volume limit), the oscillatory frequencies are indexed by $k\in{\mathbb R}^3$, and $\omega_k$, $g_k$, $a^*_k$ and $a_k$ become functions $\omega(k)$, $g(k)$, $a^*(k)$, $a(k)$ with $[a(k),a^*(\ell)] = \delta(k-\ell)$ (Dirac function).  In the continuous mode limit, the reservoir Hamiltonian \eqref{3} and field operator \eqref{4}  are
\begin{align}
	\label{5}
	H_\r &= \int_{{\mathbb R}^3} \omega(k) a^*(k)a(k)d^3k,\nonumber\\
	\varphi(g) &= \frac{1}{\sqrt 2}\int_{{\mathbb R}^3} \big(g(k) a^*(k) +{\rm h.c.}\big) d^3k.
\end{align}
The Hilbert space on which the operators \eqref{5} act is the Bosonic Fock space over the single particle wave function space $L^2({\mathbb R}^3,d^3k)$ (momentum representation), \blue{
\begin{equation}
	\label{36.f}
	{\mathcal F} = \oplus_{n\ge 0} \, L_{\rm sym}^2({\mathbb R}^{3n}, d^{3n}k), 
\end{equation}
where the subscript sym refers to symmetric functions  (Bosons) and the summand with $n=0$ is interpreted to be $\mathbb C$.} 

It is customary in the physics literature to carry out calculations for discrete modes (\eqref{3}, \eqref{4}) and take the continuous limit in quantities of interest at the end. However, it might be advantageous to start off directly with the continuous model, because then one can attack the dynamical problem by spectral analysis of the Hamiltonian, using that continuous spectrum is associated with scattering effects and irreversibility. This is the approach we take here. A (minor) trade off is that in the continuous mode models, defining the equilibrium state is slightly more complicated: while the operator $\e^{-\beta H_\r}$ has a finite trace for \eqref{3} this is not the case when $H_\r$ has continuous spectrum, \eqref{5}. The notion of reservoir equilibrium density matrix $\rho_{\r,\beta}\propto \e^{-\beta H_\r}$ has therefore to be replaced by that of a state (normalized linear functional) $\omega_{\r,\beta}$ on reservoir observables. The latter is obtained  by taking the thermodynamic limit of the discrete mode model and is determined entirely by its two point function ($k,l\in{\mathbb R}^3$)
\begin{equation}
	\label{31}
	\omega_{\r,\beta}\big( a^*(k) a(l) \big) = \frac{\delta(k-l)}{\e^{\beta \omega(k)}-1 }.
\end{equation}
Averages of general reservoir observables are found using Wick's theorem (quasi free, or Gaussian state).  We explain this in Section \ref{sect2}. The analysis presented here can be carried out for more general states, where the right side of \eqref{31} is replaced by $\mu(k) \delta(k-l)$ for general functions $\mu(k)>0$, see {\em e.g.} Section 4.3 of \cite{Mlnotes}. Having in mind spectral methods, as mentioned above, it will be useful to take a purification of reservoir state, {\em i.e.}, to describe $\omega_{\r,\beta}$ by a vector state in a (new) Hilbert space.

In this paper, it is understood that the continuous mode limit is performed and all statements are given for continuous models. In other words, we consider Hamiltonians \eqref{1} with $H_\r$ and $\varphi(g)$ given in \eqref{5}. 

\blue{
\subsection{Initial states}

We consider initial states  belonging to the {\em folium} of the reference state $\omega_{\rm ref}=\omega_\s\otimes\omega_{\r,\beta}$, where $\omega_\s( \cdot ) = \frac1N{\rm tr}( \cdot )$ is the trace state on the system and  $\omega_{\r,\beta}$ is the reservoir equilibrium state in the thermodynamic limit, characterized by \eqref{31}.\footnote{\blue{By definition (see for instance \cite{Haag}), a state $\omega$ belongs to the folium of a state $\omega_{\rm ref}$ if $\omega$ is represented by a density matrix in the Hilbert space of $\omega_{\rm ref}$. More precisely, let $({\mathcal H}, \pi,\Omega)$ be the Gelfand-Naimark-Segal (GNS) representation of $\omega_{\rm ref}$, {\em i.e.}, $\omega_{\rm ref}(A) = \langle \Omega, \pi(A)\Omega\rangle$ for all observables $A$ and where the inner product is that of $\mathcal H$. Then the folium is the collection of all states $\omega$ such that $\omega(A) = {\rm tr}_{\mathcal H} \big( \varrho \pi(A)\big)$, where $\varrho$ is any density matrix on $\mathcal H$. 
}}   These states are also called normal states with respect to $\omega_{\rm ref}$  \cite{BR, Haag}. The folium contains the states which are spatially asymptotically close to equilibrium. To explain what this means, denote by $\tau_x$, $x\in{\mathbb R}^3$, the translation automorphism group acting on reservoir observables, so that if $A_{\r,\rm loc}$ is a reservoir observable (for instance a polynomial of creation and annihilation operators) supported in a bounded region ${\mathcal R}\subset {\mathbb R}^3$, then $\tau_x(A_{\r,\rm loc})$ is its translate, supported on ${\mathcal R} +x$. Now let $\omega= \omega_\s\otimes\omega_\r$ be a disentangled state in the folium of $\omega_{\rm ref}$. Then we have $\lim_{|x|\rightarrow\infty} \omega(\bbbone_\s\otimes \tau_x(A_{\r,\rm loc}))=\omega_{\r,\beta}(A_{\r,\rm loc})$.\footnote{\blue{$\omega_\r$ is a convex combination of states of the form $\langle \pi(X) \Omega_{\r,\beta}, \pi(\cdot) \pi(X) \Omega_{\r,\beta}\rangle$, where $X$ is a local (or quasilocal) unitary operator and $\Omega_{\r,\beta}$ is the GNS vector representing $\omega_{\r,\beta}$. Due to (quasi-) locality  we have $\lim_{|x|\rightarrow\infty} [\tau_x(A_{\r,\rm loc}), X]=0$  (commutator) which implies the statement.}} (For convex combinations of such states the argument is similar.) This is the meaning of the asymptotical equilibrium property.

The reason for our choice of initial states is easily understood: our methods rely on representing the initial state on a Hilbert space and relate the dynamics to spectral properties of the generator of dynamics. All states represented in the {\em same} Hilbert space can then be dealt with on the same footing.}  Notice that within this folium, the initial system-reservoir states are allowed to be {\em entangled}. We explain this point below in Section \ref{sect2}, and \eqref{83} is our fundamental result for the dynamics, equally valid for entangled and product initial states. The dynamics for non-factorized initial states in the van Hove (weak coupling regime) was analyzed in \cite{Tetal,Yetal} (see also the references therein) and we will address the detailed analysis of our results on the dynamics of entangled states elsewhere. 
\bigskip

The main goal of Sections \ref{subs1}-\ref{subs3} is to make a link with the usual setup and results in open system theory, where the system dynamics is given by a propagator $V_t$. The latter is well defined for disentangled initial states of the form $\rho_\s\otimes\rho_{\r,\beta}$, where $\rho_{\r,\beta}$ is the equilibrium state (in the thermodynamic  limit) and $\rho_\s$ is an arbitrary system state. (Strictly speaking, $\rho_{\r,\beta}$ here is the density matrix representing $\omega_{\r,\beta}$ in the purification Hilbert space -- this point is explained in detail in Section \ref{sect2}.) The system dynamics is described by the {\em reduced system density matrix} 
\begin{equation}
	\label{32}
	\rho_\s(t) = {\rm tr}_\r\, \e^{-\i t H} (\rho_\s\otimes\rho_{\r,\beta}) \,  \e^{\i t H},
\end{equation}
where ${\rm tr}_\r$ is the partial trace over the reservoir degrees of freedom. The relation \eqref{32} defines a linear map  on system density matrices, called the {\em dynamical map $V_t$},  by
\begin{equation}
	\label{7}
	\rho_\s\mapsto V_t\rho_\s \equiv \rho_\s(t),
\end{equation}
\blue{and where $\equiv$ denotes a definition.} 
Equivalently, one can introduce the Heisenberg dynamics $t\mapsto \alpha_t A$ of system observables $A$ (hermitian matrices acting on the system), by setting
\begin{equation}
	\label{8}
	{\rm tr}_\s \,(V_t\rho_\s) A = {\rm tr}_\s\, \rho_\s (\alpha_t A).
\end{equation}
It is well known (and a source of great difficulty in theory and applications) that the map $t\mapsto V_t$ is not a group in $t$, namely $V_{t+s}\neq V_t\circ V_s$. Of course, for $\lambda=0$, $V_t\rho_\s = \e^{-\i t H_\s} \rho_\s \e^{\i t H_\s}$ {\em does} have the group property, but when the system interacts with the reservoir ($\lambda\neq 0$), correlations between the two are built up and the group property is destroyed. Still, being the reduction of a unitary dynamics of a bigger physical system (namely, the system plus the reservoir), the reduced dynamics $V_t$ has a special structure. Indeed, for each $t$ fixed, $V_t$ is a {\em completely positive, trace preserving map}, for short, $V_t$ is CPT \footnote{A map $V$ acting on ${\mathcal B}({\mathcal H})$, the bounded operators on a Hilbert space $\mathcal H$, is called CPT if (i) for all $\rho\in{\mathcal B}({\mathcal H})$ having finite trace, ${\rm tr} V\rho = {\rm tr}\rho$ (trace preserving) and (ii) $V\otimes{\bbbone }$ is positivity preserving on the space of operators ${\mathcal B}({\mathcal H})\otimes{\mathcal B}({\mathbb C}^K)$, for all $K\ge 1$ (complete positivity). Positivity preserving in turn means that if $X$ is a bounded non-negative operator acting on ${\mathcal H}\otimes{\mathbb C}^K$ (having non-negative spectrum only), then $(V\otimes{\bbbone})X$ is a bounded non-negative operator acting on ${\mathcal H}\otimes{\mathbb C}^K$. If $V$ is completely positive then it is positivity preserving, but the converse is not true. For instance, consider two qubits and take $V$ to the partial transpose operator. This is a positivity preserving map but it is not CP. Indeed the positive partial transpose (PPT) criterion to check for entanglement in quantum information theory is based on the fact that the partial transpose is not CP.}.  Using \eqref{8} it is not difficult to understand that, for any $t$ fixed, $V_t$  is CPT if and only if $\alpha_t$ is completely positive and identity preserving ($\alpha_t{\bbbone} = \bbbone$).

\blue{
\subsection{Importance of the group property}
}
If the group property $V_{t+s}=V_t\circ V_s$ is satisfied, then there is a generator $\mathcal L$, a linear operator acting on density matrices, such that $V_t =\e^{t{\mathcal L}}$. The open system dynamics is entirely determined by the spectral data (eigenvalues and eigenvectors) of $\mathcal L$. Assume for the moment that one can show a spectral representation
\begin{equation}
	\label{11}
	\e^{t{\mathcal L}} = \sum_j \e^{\i t \epsilon_j} P_j,
\end{equation}
where $\i \epsilon_j$ are the eigenvalues of $\mathcal L$ and $P_j$ the corresponding eigenprojections.\blue{\footnote{\blue{This diagonalization property is {\em assumed} here. It is satisfied if all eigenvalues are simple, for example. Our method works equally well in case $\mathcal L$ has Jordan blocks but we do not address this point here.}}}  All dynamical information is then contained in the $\epsilon_j$ and $P_j$. Namely, the $\epsilon_j$ with ${\rm Im} \,\epsilon_j>0$ drive irreversible decay ($t>0$), with decay rates ${\rm Im}\, \epsilon_j$ and the associated $P_j$ determine the decay directions in state space. Stationary states are in the range of the projections $P_j$ with $j$ such that $\epsilon_j=0$.
\bigskip

\blue{
\subsection{Importance of complete positivity}
}
Suppose you have a bipartite system $AB$ in an entangled initial state $\rho_{AB}$. Suppose that the  subsystem $B$ evolves independently, according to its own unitary dynamics $U_t$ (generated by a Hamiltonian $H_B$) and that the dynamics of subsystem $A$ is given by $V_t$ (emerging for instance by interaction with a reservoir). The state of $AB$ at time $t$ is then $\rho_{AB}(t) = (V_t\otimes U_t)\rho_{AB}(0)$. This state is guaranteed to be a density matrix only because $V_t$ is completely positive. (If $V_t$ was not completely positive, then one could find an initial density matrix $\rho_{AB}(0)$ for which $\rho_{AB}(t)$ would have some negative eigenvalues!) On the mathematical side, complete positivity of a map $V$ is equivalent with $V$ having a Kraus representation, which is again equivalent with $V$ being the reduction of a unitary map acting on a bigger system (adding an ancilla reservoir system). We refer to \cite{BF,AL,BP,CP} for more detail about this.
\bigskip

\blue{
\subsection{Markovian approximation in the van Hove weak coupling regime}
}
 Intuitively, if the reservoir dynamics is very fast, maybe if local disturbances of the reservoir state are quickly propagated far away (short lived reservoir memory), and if the system-reservoir interaction is not too large, then the back reaction from the reservoir onto the system might be minor. In this situation, one expects the group property to hold for $t\mapsto V_t$. Quantifying this idea is an important problem, leading to the {\em Markovian approximation}. The challenge is to show the validity of a Markovian approximation
\begin{equation}
	\label{10}
	V_t = \e^{t {\mathcal L}} +R(t,\lambda)
\end{equation}
and to find a parameter regime in which the remainder term  $R(t,\lambda)$ is small. When the remainder is squarely neglected, $V_t=\e^{t{\mathcal L}}$ is the integrated version of the differential equation $\frac{d}{dt} V_t = {\mathcal L} V_t$, or as per  \eqref{7}, $\frac{d}{dt}\rho_\s(t) = {\mathcal L}\rho_\s(t)$, which is called the {\em Markovian master equation} for the system density matrix $\rho_\s(t)$.  It is a difficult problem to find quantitative and controlled (not heuristic) bounds on the remainder $R(t,\lambda)$ in \eqref{10}. There is one rigorous approach, called the {\em van Hove-}, or {\em weak coupling limit}. It states that for all $a>0$,
\begin{equation}
	\label{12}
	\lim_{\lambda\rightarrow 0} \sup_{0\le \lambda^2 t<a} \big\| V_t - \e^{t ({\mathcal L}_\s+\lambda^2 K)}\big\|=0.
\end{equation}
Here, ${\mathcal L}_\s$ and $K$ are commuting operators acting on system density matrices and for each $t$ fixed, $e^{t({\mathcal L}_\s+\lambda^2 K)}$ is CPT.\footnote{Which norm $\|\cdot\|$ we take in \eqref{12} is not too important here, as we assume that the system Hilbert space has finite dimension and so all norms are equivalent.} The operator ${\mathcal L}_\s = -\i [H_\s,\cdot]$ generates the free system dynamics (no interaction) and $K$ is a (lowest order) correction term, encoding coupling effects. The $\lambda^2 t$ scaling was used in \cite{VH} and later analyzed with mathematical rigour in \cite{D, D1}. The literature on the weak coupling regime and Markovian master equations is huge and growing. It has important applications not only to physics and mathematics, but also to chemistry, biology and the quantum information sciences \cite{RPHP, HR, Mohseni, BFFP}. It is worthwhile to note that many different (heuristic) approximations and candidates for generators have been proposed over time, often violating the CPT requirement, with the Davies generator ${\mathcal L}_\s+\lambda^2K$ above emerging as the correct one \cite{DS,SB}. However, from a practical (numerical) perspective and in combination with
other methods, different generators might be more advantageous and might be able to describe
specific phenomena in more detail \cite{SSSE}.

The relation \eqref{12} is the same as \eqref{10} with ${\mathcal L}={\mathcal L}_\s+\lambda^2 K$ and \eqref{12}  says 
\begin{equation}
	\label{13}
	\lim_{\lambda\rightarrow 0} \sup_{0\le \lambda^2 t<a} \big\| R(t,\lambda)\big\|=0.
\end{equation} 
The shortcoming of \eqref{12}, \eqref{13} is that only times up to $t\approx a/\lambda^2$ are resolved by the Markovian approximation. Beyond that time scale, $\e^{t ({\mathcal L}_\s+\lambda^2 K)}$ is not guaranteed to be accurate (the remainder may not be small). Of course, $a$ is arbitrary, so in principle one can consider large times -- but the bigger one takes $a$, the smaller $\lambda$ has to be in order to make the remainder smaller than a given accuracy. (In other words, the speed of convergence in \eqref{13} depends on $a$). Another way of saying this is that, when considering $t\rightarrow\infty$ one has to take at the same time $\lambda\rightarrow 0$ in such a way that $\lambda^2 t$ stays bounded ($<a$), in order to be sure that the Markovian approximation is valid. This is called the van Hove weak coupling regime.

One of our main results is to remove the condition that $\lambda^2 t$ needs to be bounded. We show the accuracy of the Markovian approximation {\em for all times $t\ge 0$}. 
\bigskip

\blue{
\subsection{Regularity assumption on the form factor and decay of reservoir correlations}
} 
The symmetrized correlation function is defined as \blue{
\begin{equation}
	\label{m7}
C_\beta(t) = {\rm Re}\, \omega_{\r,\beta}\big( \varphi(g) \e^{\i t H_\r}\varphi(g) \e^{-\i t H_\r} \big)  = {\rm Re}\, \omega_{\r,\beta}\big(  \varphi(g)  \varphi(\e^{\i\omega t} g)\big) ,
\end{equation} 
where $g$ is the form factor in the interaction \eqref{1} and $\omega_{\r,\beta}$ is the reservoir thermal equilibrium state \eqref{31}. The free reservoir dynamics is characterized by the Bogoliubov transformation $g(k)\mapsto \e^{\i\omega(k)t}g(k)$ (see also \eqref{5}). The resonance theory we develop requires a regularity condition on the function $g$. To state it, define the complex valued function 
\begin{equation}
	\label{m1.1}
	g_\beta(u,\Sigma) = \sqrt{\frac{u}{1-\e^{-\beta u}}}\,  |u|^{1/2} \left\{
	\begin{array}{ll}
		g(u,\Sigma) & u\ge 0\\
		- \e^{\i\alpha}\bar g(-u,\Sigma) & u<0
	\end{array}
	\right.,
\end{equation}
where $g\blue{(r,\Sigma)}$ is the form factor $g$ expressed in spherical coordinates, $r\ge 0$ and $\Sigma\in S^2$.  In  \eqref{m1.1}, $u\in\mathbb R$, so $g_\beta$ is a function of ${\mathbb R}\times S^2$, while the original $g$ is a function of ${\mathbb R}_+\times S^2={\mathbb R}^3$. The phase $\alpha\in\mathbb R$ can be chosen arbitrarily. 
\medskip

{\bf We assume the following condition.}
\smallskip

\begin{itemize}
	\item[{\bf (A)}] For $\theta \in\mathbb R$, set $(T_\theta g_\beta)(u,\Sigma) = g_\beta(u-\theta,\Sigma)$. There exists a $\theta_0>0$ such that $\theta \mapsto T_\theta g_\beta$ has an analytic extension (as a function from $\mathbb R$ to $L^2({\mathbb R}\times S^2)$) to $0<{\rm Im }\theta<\theta_0$ which is continuous at ${\rm Im}\theta\rightarrow 0_+$.
\end{itemize} 
Note that the $\theta$ in condition (A) is not an angle, rather it is a parameter of {\em t}ranslation. A condition which is less technical and which implies (A) is the following. 
\begin{itemize}
\item[{\bf (H)}] Suppose that for some $\theta_0>0$, the function $u\mapsto g_\beta(u,\Sigma)$ extends to an analytic function for complex values of $u$ belonging to the strip $S_{\theta_0}=\{ z\in{\mathbb C}\ :\ -\theta_0 < {\rm Im} z <\theta_0\}$, for all $\Sigma \in S^2$, such that 
\begin{equation}
	\label{bound}
c_0=\sup_{-\theta_0< v<\theta_0}\ \int_{\mathbb R} du \int_{S^2} d\Sigma\  \big| g_\beta(u+\i v,\Sigma)\big|^2 <\infty.
\end{equation}
Functions $g_\beta$ having this property belong to the Hardy class on the strip $S_{\theta_0}$. Condition (H) implies condition (A). 
\end{itemize} 
\medskip

\noindent
{\bf Discussion of assumptions (A) and (H)}
\begin{itemize}

\item[(1)] The function $g$ has to behave appropriately in the infrared regime so that the parts of \eqref{m1.1}  fit nicely together at $u = 0$, to allow for an
analytic continuation. The square root in \eqref{m1.1} must be analytic as well,
which implies the condition $\theta_0< 2\pi/\beta$. This means that we have to consider strictly positive temperature $T=1/\beta >0$. Furthermore, our method requires an upper bound $\lambda^2 \le c \theta_0$ (some fixed $c$), see {\em e.g.} Figure 1 below, so we need the condition $\lambda^2 \le c T$ to hold. 
A family of form factors $g$ satisfying condition (A) is given by
$$
g(r,\Sigma) = r^p \e^{-r^2} g_1(\Sigma), \quad \mbox{ with $p=-1/2+n$, $n=0,1,2,\ldots$},
$$
and where $g_1$ is an arbitrary function of the angle $\Sigma\in S^2$ satisfying $g_1(\Sigma)=\e^{\i\alpha'}\bar g_1(\Sigma)$ for an arbitrary phase $\alpha'\in\mathbb R$. As an example, suppose $g(r,\Sigma) = r^{-1/2} \e^{-r^2}$. Then we chose $\alpha=\pi$ in \eqref{m1.1} and get $g_\beta(u,\Sigma) =\e^{-u^2} \sqrt{\frac{u}{1-\e^{-\beta u}}}$ which satisfies (H), hence (A). If $g(r,\Sigma) = r^{1/2} \e^{-r^2}$, then we choose $\alpha=0$ in \eqref{m1.1} and obtain $g_\beta(u,\Sigma) =u \e^{-u^2} \sqrt{\frac{u}{1-\e^{-\beta u}}}$, which again satisfies (H) and (A).

\item[(2)] Assumption (A) guarantees that the simplest version of spectral deformation techniques is applicable (namely, spectral translation). The reservoir correlation function \eqref{m7} can be written as  
$$
C_\beta(t) = \int_{\mathbb R} d u\, \e^{\i u t}\int_{S^2} d\Sigma\,  |g_\beta(u,\Sigma)|^2
$$
(this is a direct calculation, see also \cite{MSoB}, Appendix A)  and (A)  implies exponential decay of the correlation function. Indeed, let $\theta'<\theta_0$, then 
$$
C_\beta(t) = \e^{-\theta' t} \int_{\mathbb R} d u\, \e^{\i (u-\i\theta') t}\int_{S^2} d\Sigma\,  |g_\beta(u,\Sigma)|^2.
$$
By a change of the variable $u$, the integral is over the function $|g_\beta(u+\i\theta',\Sigma)|^2$, which is finite and therefore, $C_\beta(t)\le {\rm const.} \e^{-\theta' t}$. In this paper, we thus assume that the reservoir correlation decay exponentially. An extension of the theory to polynomially decaying correlations is possible and is planned, see point (4) below.

\item[(3)] The property of return to equilibrium for open systems, that is, that initial states converge to the coupled system-reservoir equilibrium in the limit of large times (see also \eqref{rte}),  was shown under milder regularity conditions on $g$ by using a combination of spectral dilation and translation and a rather involved renormalization group analysis in \cite{BFS}. However, those techniques have never been used to address the much more detailed information on the reduced system dynamics as we do here in Sections \ref{subs1}-\ref{subs3}. 

\item[(4)] Another approach is to replace spectral deformation theory by an infinitesimal version of it, called  {\em Mourre theory}. Here one can significantly weaken the regularity requirements on $g$, replacing analyticity by just real differentiability. This implies that $C_\beta(t)$ decays polynomially in time only. This technically more demanding route was taken in \cite{KoMeDyn} to find the reduced dynamics of the system modulo an error $\propto 1/t$ for large times. The fact that weaker regularity of $g$ leads to polynomial decay of errors (while analyticity implies exponential decay in our results here, Sections \ref{subs1} - \ref{subs3}) might not be surprising. Tthe disadvantage of \cite{KoMeDyn} is that the remainder is not shown to be small in $\lambda$. An extension of the techniques of \cite{KoMeDyn} to show this smallness is planned. It would yield results presented here for a much larger class of form factors $g$, so a less stringent regularity condition can be traded off for polynomially decaying remainder terms. 

\item[(5)] The fact that (H) implies (A) can be seen as follows. First note that weak analyticity is equivalent with strong analyticity \cite{RS1} (Theorem VI.4). This means we only have to show that for each $f\in L^2=L^2({\mathbb R\times S^2, du\times d\Sigma})$, the complex valued function  $\theta\mapsto \langle f, g_\beta(\cdot -\theta)\rangle = \int_{\mathbb R} du \int_{S^2} d\Sigma\  \overline{{f(u,\Sigma)}} g_\beta(u-\theta,\Sigma)$ extends to complex values of $\theta$ with $0<{\rm Im \theta}<\theta_0$ which is continuous at the real line. First take $f$ with compact support in the variable $u$. Then you may interchange $d/d\theta$ with the integrals,
$$
\tfrac{d}{d\theta}  \langle f, g_\beta(\cdot -\theta)\rangle = \int_{\mathbb R} du \int_{S^2} d\Sigma\   \overline{{f(u,\Sigma)}} \, \tfrac{d}{d\theta} g_\beta(u-\theta,\Sigma)
$$
and it is now clear that $\theta\mapsto \langle f, g_\beta(\cdot  -\theta)\rangle$ is analytic in $S_{\theta_0}$  due to condition (H). To show analyticity for an arbitrary $f$, take a sequence of functions $f_n\in L^2$ with compact support in $u$ satisfying $\|f_n-f\|_{L^2}\rightarrow 0$. Each $F_n(\theta) = \langle f_n, g_\beta(\cdot -\theta)\rangle$ is analytic by the previous argument. Moreover, since $|F_n(\theta) - \langle f, g_\beta(\cdot-\theta)\rangle| \le c_0 \|f_n-f\|_{L^2}$, where $c_0$ is given in \eqref{bound}, we have that   $F_n(\theta)$ converges to $\langle f, g_\beta(\cdot -\theta)\rangle$ uniformly in $\theta\in S_{\theta_0}$. Hence the limit function is also analytic.
\end{itemize}

}

\subsection{Result 1: Resonance expansion of the dynamics}

\label{subs1} The resonance theory is a mathematically rigorous approach for the analysis of the evolution of the system-reservoir complex. It does not only describe the dynamics of the system state or observables, but also that of the reservoir. Here we explain the results on the system Schr\"odinger dynamics. To state our results in terms of the dynamical map $V_t$, we assume that the initial system-reservoir state is disentangled, of the form \eqref{32} for $t=0$. (The result for general initial states is given in \eqref{83}.)

We show that if $|\lambda|\le \lambda_0$ (for some $\lambda_0>0$), then for all times $t\ge 0$,
\begin{equation}
	\label{14}
	\big\| V_t -W_t -  \rho_{\s,\beta,\lambda} \,\langle {\rm tr}| \, \big\| \le C \lambda^2 \e^{-\gamma(\lambda) t}. 
\end{equation}
The constant $C<\infty$ is independent of $\lambda$, $t$ and $\gamma(\lambda)\ge 0$ does not depend on $t$.   In \eqref{14}, $\,\langle {\rm tr}| \, $ is the linear functional $\rho\mapsto {\rm tr}(\rho )=1$. Moreover, $\rho_{\s,\beta,\lambda}$ is the effective system equilibrium state, obtained by taking the full, coupled system-reservoir equilibrium state (relative to $H$, \eqref{1}) and tracing out the reservoir degrees of freedom.  $W_t$ is a linear map on system states (density matrices), describing how, and if, the system approaches the equilibrium $\rho_{\s,\beta,\lambda}$.
It  has an expansion of the type \eqref{11},
\begin{equation}
	\label{15}
	W_t = \sum_j \e^{\i t\epsilon_j(\lambda)}  {\mathcal P}_j,
\end{equation} 
where the ${\mathcal  P}_j$ are $\lambda$-independent projection operators (acting on system density matrices). They satisfy 
\begin{equation}
	\label{15.1}
	{\mathcal P}_j {\mathcal P}_k = \delta_{j,k} {\mathcal P}_j\quad \mbox{and}\quad \sum_j {\mathcal P}_j =W_0 = \bbbone -\rho_{\s,\beta,0} \,\langle {\rm tr}| ,
\end{equation}
where $\rho_{\s,\beta,0}= \e^{-\beta H_\s}/{\rm tr}(\e^{-\beta H_\s})$ is the (uncoupled) system equilibrium state. The $\epsilon_j(\lambda)\in {\mathbb C}$ are analytic in $\lambda$ at the origin,
\begin{equation}
	\label{23}
	\epsilon_j(\lambda)=\epsilon_j^{(0)} +\lambda^2 \epsilon_j^{(2)} +O(\lambda^4)
\end{equation}
and $\epsilon_j^{(0)}$ are differences of eigenvalues of $H_\s$ (Bohr energies).  It is clear from \eqref{15} and the properties of the ${\mathcal P}_j$ that 
\begin{equation}
	\label{23.1}
	W_{t+s} = W_t\circ W_s.
\end{equation}  
Symmetries or degeneracies in the spectrum of $H_\s$ can cause some of the $\epsilon_j(\lambda)$ to vanish (or to be real). In this case, the associated ${\mathcal P}_j$ project onto additional stationary states, other than $\rho_{\s,\beta,\lambda}$. However, generically, in the absence of symmetries and degeneracies, one has ${\rm Im}\epsilon_j(\lambda) >0$ for all $j$ (for small, nonzero $\lambda$). Then all terms in \eqref{15} decay in time, the $j$th one at the rate ${\rm Im}\epsilon_j(\lambda)$. Denoting by $2\ell_j$ the order of the zero of ${\rm Im}\epsilon_j(\lambda)$ at the origin, {\em i.e.}, ${\rm Im}\epsilon_j \propto \lambda^{2\ell_j}$ to leading order in $\lambda$, we see that $W_t$ is a sum of terms decaying at (possibly different) rates $\lambda^{2\ell_j}$. The slowest decay rate is
\begin{equation}
	\label{77}
	\gamma(\lambda) = \min_j {\rm Im }\epsilon_j(\lambda) \ge 0
\end{equation}
and coincides with that of the remainder in \eqref{14}. Note, however, the additional factor $\lambda^2$ on the right side of \eqref{14}. The result \eqref{14} can be expressed as
\begin{equation}
	\label{16}
	V_t\rho = \rho_{\s,\beta,\lambda} + W_t\rho +O\big( \lambda^2 \e^{-\gamma(\lambda)t}\big)
\end{equation}
for any density matrix $\rho$, with an error term which is (quadratically) {\em small in $\lambda$  for all times}, and which also {\em decays to zero exponentially quickly in time}.

\subsection{Result 2: Approximation of the dynamics by a  CPT semigroup  for all times} 
In applications it is often observed that the imaginary parts of all the $\epsilon_j(\lambda)$  are strictly positive already to second order in $\lambda$ (see \eqref{23}), {\em i.e.}, that  
\begin{equation}
	\label{22}
	\gamma_{\rm FGR}\equiv \min_j  {\rm Im}\, \epsilon_j^{(2)}  >0. 
\end{equation}
If \eqref{22} is satisfied we say that the Fermi Golden Rule Condition holds \cite{Alicki1,BFS,JP,MSB1,MSB2}. In this situation,  $W_t$ contains the single characteristic time scale $\lambda^{-2}$.  We assume \eqref{22} now. Retaining only the leading terms of $W_t$ and $\rho_{\s,\beta,\lambda}$ on the left side of \eqref{14}, namely
\begin{equation}
	\label{25}
	\epsilon_j(\lambda) \approx \epsilon_j^{(0)} +\lambda^2 \epsilon_j^{(2)},   \ \ \rho_{\s, \beta,\lambda}  \approx \rho_{\s,\beta,0}=\frac{\e^{-\beta H_\s}}{{\rm tr}\, \e^{-\beta H_\s}},
\end{equation}
we can show the following result. There is a $\lambda_0>0$ such that if $|\lambda| \le \lambda_0$, then for all $t\ge 0$,
\begin{equation}
	\label{24}
	\big\| V_t -\e^{t({\mathcal L}_\s+\lambda^2 K)}\big\| \le C \lambda^2.
\end{equation}
Here, ${\mathcal L}_\s = -\i [H_\s,\cdot]$ (commutator) and $K$ are commuting operators acting on system density matrices, and $K$ is constructed entirely in terms of $\epsilon_j^{(2)}$ and ${\mathcal P}_j$.  Moreover, {\em $\e^{t({\mathcal L}_\s +\lambda^2K)}$ is a CPT semigroup} satisfying
\begin{equation}
	\label{26}
	\e^{t({\mathcal L}_\s+\lambda^2 K)}\rho_{\s,\beta,0} = \rho_{\s,\beta,0}.
\end{equation}
It is the same semigroup as the one in the weak coupling (van Hove) result \eqref{12}. 
In passing from \eqref{15} to \eqref{24} we have gained the CPT and semigroup properties of the approximation, but we have traded it for a worse error estimate. Namely, the approximation \eqref{24} is still $O(\lambda^2)$ for all $t\ge 0$, but it does not decay to zero for large times, as it did in \eqref{14}. The inequality \eqref{24} proves that the Markovian approximation, implemented by a CPT semigroup, is valid for all times $t\ge0$. It can be phrased as 
\begin{equation}
	\label{13.01}
\sup_{t\ge 0} \big\| V_t-\e^{t({\mathcal L}_\s+\lambda^2 K)}\big\|\le C\lambda^2.
\end{equation}
This is a significant improvement of the weak coupling result \eqref{12}. 

The generator $K$ can be obtained by perturbation theory or by the relation 
\begin{equation}
	\label{-1}
	\lim_{\lambda\rightarrow 0}\ V_{\frac{\tau}{\lambda^2}} \circ\e^{-\frac{\tau}{\lambda^2} {\mathcal L}_\s} = \e^{\tau K},\qquad \tau\ge 0,
\end{equation}
which identifies it as the Davies generator (the same $K$ as in  \eqref{12}), \cite{D,D1,AL,BP,CP,FP}. It can be calculated explicitly, see the Appendix \ref{appendixA}.

\subsection{Result 3: Approximation of the  dynamics by an asymptotically exact CPT semigroup } 
\label{subs3}
The origin of the loss of time decay in the remainder, when passing from \eqref{14} to \eqref{24} as described in the previous section, comes from replacing $\rho_{\s,\beta,\lambda}$ by $\rho_{\s,\beta,0}$ (see \eqref{25}). We recall that $\rho_{\s,\beta,\lambda}$ is the restriction to the system of the full, coupled system-reservoir equilibrium state. This replacement unavoidably introduces an error of $O(\lambda^2)$ for large times, as the true final ($t\rightarrow\infty$) system state is $\rho_{\s,\beta,\lambda}$, while the one predicted by the approximation is $\rho_{\s,\beta,0}$, differing from the true one by $O(\lambda^2)$. Above, this replacement was necessary in order to incorporate the final state  into the approximate dynamical group, as an element in the kernel of the generator ${\mathcal L}_\s+\lambda^2K$, see \eqref{26}. To avoid the approximation of $\rho_{\s,\beta,\lambda}$, we might try to modify the generator into a new one, $M(\lambda)$,  by adding supplementary terms of all orders in $\lambda$, as to make the full $\rho_{\s,\beta,\lambda}$ an invariant state. This is the result we explain now, and in this result we restore the time decay of the remainder (obtaining thus an asymptotically exact approximation). 

We introduce a  renormalization, $\widetilde H_\s(\lambda)$,  of the system Hamiltonian, satisfying
\begin{equation}
	\label{27}
	\frac{\e^{-\beta \widetilde H_\s(\lambda)}}{{\rm tr} \e^{-\beta \widetilde H_\s(\lambda)}} = \rho_{\s,\beta,\lambda}.
\end{equation}
By carrying out the resonance theory leading to the results of Subsection \ref{subs1}, but now with this renormalized reference state \eqref{27}, the CPT semigroup approximating the true dynamics $V_t$ turns out to be $\e^{t(\widetilde{\mathcal L}_\s+\lambda^2\widetilde K)}$, with {\em  $\lambda$ dependent operators $\widetilde {\mathcal L}_\s$ and $\widetilde K$}. The crucial point is that $\e^{t(\widetilde{\mathcal L}_\s+\lambda^2\widetilde K)}\rho_{\s,\beta,\lambda}=\rho_{\s,\beta,\lambda}$, which replaces the property \eqref{26} in the previous argument and allows us to obtain a remainder which decays to zero for large times. We show the following. 

Suppose that the Fermi Golden Rule Condition $\gamma_{\rm FGR}>0$ is satisfied (c.f. \eqref{22}). Then there is a $\lambda_0>0$ such that for $|\lambda|< \lambda_0$, and all times $t\ge 0$,
\begin{equation}
	\label{28}
	\big\|  V_t-\e^{t M(\lambda)}\big\| \le C \big(|\lambda|+\lambda^2 t\big) \, \e^{-\lambda^2\gamma_{\rm FGR} \,t\,  (1+O(\lambda^2))}.
\end{equation}
Here, $\e^{t M(\lambda)}$ is a CPT semigroup with a generator $M(\lambda)$  analytic in $\lambda$, containing all orders of $\lambda$. Its Taylor series can be calculated by perturbation theory. The result \eqref{28} shows that we can construct a CPT semigroup which approximates the true dynamics and which is asymptotically exact, meaning that $\lim_{t\rightarrow\infty} (V_t  -  \e^{t M(\lambda)})=0$. Note, however, that for $t\sim 1/\lambda^2$, the right hand side of \eqref{28} is not small. Still, for times $t> 1/( \lambda^2 \gamma_{\rm FGR})$ the remainder becomes negligible.

\blue{
We obtain a better result for the dynamics of observables which commute with $H_\s$ (or, for the populations of the system density matrix). Namely, we show that there is a $\lambda_0>0$ such that for $|\lambda| <  \lambda_0$,
\begin{eqnarray}
\big\| V_t\circ V^\s_{-t} - \e^{t\lambda^2 M_{\rm d}(\lambda)} - (\bbbone_\s - V^\s_{-t})  \rho_{\s,\beta,\lambda} \  \langle {\rm tr}|\  \big\|\le C (\lambda+\lambda^4t)  \e^{- \lambda^2 t (\gamma_{\rm FGR} + O(\lambda^2))}.
	\label{106.4}
\end{eqnarray}
Here,  $V^\s_{-t}$ is the free system dynamics, $V^\s_{-t}\, \rho = \e^{\i t H_\s}\rho \e^{-\i t H_\s}$ for any system density matrix $\rho$. Moreover, $\e^{t \lambda^2 M_{\rm d}(\lambda)}$ is a CPT semigroup with a generator $M_{\rm d}(\lambda)$ analytic in $\lambda$ (d for diagonal), which is explicitly constructible by perturbation theory and satisfies $M_{\rm d}(0)=K$, the Davies generator (see \eqref{-1}). The generators $M(\lambda)$ and $M_{\rm d}(\lambda)$ are related by
\begin{equation}
	M(\lambda) = -\i [\widetilde H_\s(\lambda), \, \cdot\, ] +\lambda^2 M_{\rm d}(\lambda)
\end{equation} 
and the two operators on the right side commute.

We now show how \eqref{106.4} implies a better result than \eqref{28} for the evolution of the populations of the state $V_t\rho$, {\em i.e.}, the diagonal of the density matrix $V_t\rho$ in the energy basis of $H_\s$ (Pauli equations, see also \cite{Alicki1}). The last term on the left side in \eqref{106.4} vanishes when applied to system observables $X$ which commute with $H_\s$. Namely, let $\rho$ be a system initial state and let $X$ be such an observable. Then
\begin{equation}
\label{mn1}
{\rm tr}_\s \big( (\bbbone_\s - V^\s_{-t}) \rho_{\s,\beta,\lambda}\, \langle{\rm tr}|\, \rho\big) X={\rm tr}_\s \  \rho_{\s,\beta,\lambda}  (X - \e^{-\i t H_\s}X\e^{\i t H_\s})  =0.
\end{equation}
For an operator $A$, set
\begin{equation}
	{}[A]_{k,\ell} = \langle \phi_k, A\phi_\ell \rangle,
\end{equation}
where $\phi_k$ is the eigenvector of $H_\s$ associated to the eigenvalue $E_k$, see \eqref{2}. The population of the energy $E_k$ at time $t$ is then 
\begin{equation}
	{}[V_t\rho]_{k,k} = \langle \phi_k, (V_t\rho)\, \phi_k\rangle.
\end{equation}
Combining \eqref{106.4} and \eqref{mn1} shows that 
\begin{equation}
	{}[	V_t\rho]_{k,k} = [\e^{t \lambda^2 M_{\rm d}(\lambda)}\rho]_{k,k} 
	 + O\Big( \big(|\lambda| +\lambda^4t\big)  \e^{- \lambda^2 t (\gamma_{\rm FGR} + O(\lambda^2))}\Big),
	\label{mh4}
\end{equation}
so we have a CPT semigroup which approximates the populations to accuracy $O(\lambda)$ for all times, and on top of this, is asymptotically exact. 
}

\bigskip

\blue{{\bf Remark on the parameter dependence of constants in the error estimates and $\lambda_0$.\ } The constants $C$ in our main results \eqref{14}, \eqref{24}, \eqref{28} and \eqref{mh4} will depend on the system dimension $N$ and properties of the interaction operator $G$ and the form factor $g$. To find the dependence is in principle possible in our approach. This analysis must be carried out on a remainder that depends on all powers of $\lambda$ and we have not done this so far. We believe it would be interesting to start with a benchmark problem, say compare the approximation of the dynamics by the resonance theory to the explicit solution for the spin-boson model (or $N$ spins coupled to bosons) with energy conserving interaction. One could then  use numerical methods to compare the resonance approximation to the correct dynamics and exhibit the dependence of the difference on $N$ and on properties of the coupling function $g$ ({\rm e.g.} the ultraviolet and infrared characteristics of $g$). Similarly, one might test the validity of the resonance approximation for varying sizes of the coupling parameter $\lambda$ and find the dependence of $\lambda_0$ on model parameters.

}

\section{Mechanism of the resonance theory}
\label{sect2}

\subsection{History}
The method we develop has its origins in works using a $C^*$-dynamical system approach, pioneered in \cite{JP,BFS}. In those works, it was shown that all initial system-reservoir states $\omega$,  \blue{taken from the same class as we consider,  converge to the coupled system-reservoir equilibrium state $\omega_{\s\r,\beta,\lambda}$  in the limit of large times. More precisely, for system-reservoir observables $A$, 
\begin{equation}
	\label{rte}
\lim_{t\rightarrow\infty} \omega\big(\e^{\i t H} A \e^{-\i t H}\big) = \omega_{\s\r,\beta,\lambda}(A).
\end{equation}
In this setup, the approach to equilibrium is linked to the spectrum of the (complex deformed) Liouville operator. The spectrum of this operator consists of complex numbers and eigenvalues are called resonances. Convergence to equilibrium is implied by the fact that the Liouville operator has a simple eigenvalue at zero, the eigenvector being the equilibrium state. A spectral gap in the spectrum at the origin (when zero is an isolated resonance) makes the convergence in \eqref{rte} exponentially fast in time. This mechanism is revealed below in Section \ref{secrepdyn}.

We point out that the system-reservoir dynamics overall is Hamiltonian, governed by the unitary group $\e^{\i t H}$. So how is the relation \eqref{rte} possible? The point is that one considers only (quasi-)local observables $A$ in \eqref{rte}. To explain this, one can view the reservoir as a spatially infinitely extended reservoir of quantum particles (quantum field) in ${\mathbb R}^3$. Local observables $A$ are those made of system observables and field operators (or creation and annihilation operators) supported only at spatial points $x\in{\mathbb R}^3$ belonging to bounded sets. Quasi-local observables are limits of such observables. It becomes then intuitively clear that while the global dynamics is unitary, on local observable it is irreversible. This is just as in usual quantum theory: A single, free particle in ${\mathbb R}^3$ with Hamiltonian $-\frac{\hbar^2}{2m}\Delta_x$ having continuous spectrum will leave any bounded region as $t\rightarrow\infty$, so the average of any quasi-local observable will vanish in this limit. This happens even though the dynamics is unitary. 
}

In \cite{MSB1,MSB2} it was realized that the nonzero resonances govern the evolution of the system coherences and consequently a rigorous analysis of the dynamics of decoherence and entanglement in various physical settings became possible, see {\em e.g.}  \cite{Mdimer,MQIC}. The CPT properties and asymptotic exactness of the approximating Markovian dynamics have not been addressed until very recently. In \cite{KoMeCP} we give a short (two page) outline of a proof of the Results 1 and 2 presented in the current work. The paper \cite{KoMeCP} focuses on the construction of an asymptotically exact Markovian approximation, which is part of Result 3 of the present publication. However, there is a gap in the proof of the main result in \cite{KoMeCP}. This is explained in an erratum to \cite{KoMeCP}, where it is also announced that we can still show the result in its full strength for the dynamics of the populations of the system (but not the coherences). We give the corresponding precise statement and proof of it here in  \eqref{mh4}.

An approximate system dynamics valid for all times was constructed \cite{JP1}, using a semigroup with a generator depending on all powers of $\lambda$, but which is not asymptotically exact, and which is not shown to be CPT. In contrast, we show here that the approximation by the CPT semigroup given by the free dynamics plus the Davies generator, which is merely {\em quadratic in $\lambda$}, works for all times already. By adding higher orders in $\lambda$ to the generator, we achieve an {\em asymptotically exact CPT semigroup}. 

\blue{In this work, we only consider time independent Hamiltonians, but the resonance theory has also been applied to time dependent ones, see \cite{MS,AF, BMPS}.}

Of course, non Markovian effects play an important role in quantum physics and are heavily studied (see for instance the reviews \cite{RHP, BLPV}). A refined weak coupling limit which captures non-Markovian effects has been developed in \cite{R}.  It will be interesting to examine how our resonance theory will contribute to this line of study.

\subsection{ Purification of the initial state}

 Given any (initial) system density matrix $\rho_\s$ acting on ${\mathbb C}^N$, we take a purification, {\em i.e.}, a normalized vector $\Psi_\s  \in {\mathbb C}^N\otimes{\mathbb C}^N$ satisfying
\begin{equation}
	\label{34}
	{\rm tr}_\s \rho_\s X = \scalprod{\Psi_\s}{(X\otimes\bbbone_\s)\Psi_\s}
\end{equation}
for all system operators $X\in {\mathcal B}({\mathbb C}^N)$.\footnote{To do this explicitly, first diagonalize $\rho_\s = \sum_j p_j |\chi_j\rangle\langle\chi_j|$. \blue{Then the vector $\Psi_\s = \sum_j \sqrt{p_j} \chi_j\otimes {\cal C}\chi_j$ does the job in \eqref{34}, where $\cal C$ is any antiunitary map. Our convention is to take $\cal C$ to be the operator taking the complex conjugate of vector coordinates in the eigenbasis of $H_\s$. This purification is also known under the name of Gelfand-Naimark-Segal representation in the theory of operator algebras, and for finite dimensions in linear algebra it is called vectorization.}}
We also take a purification of the reservoir thermal equilibrium state \eqref{31}, whose associated Hilbert space is again obtained by doubling the original one, namely the Fock space $\mathcal F$, \eqref{36.f}. On ${\mathcal F}\otimes{\mathcal F}$, define the thermal annihilation operators
\begin{align}
	a_\beta (k) &= \sqrt{1+\mu(k)} \ \big( a(k)\otimes \bbbone\big)  + \sqrt{\mu(k)}\   \big( \bbbone\otimes a^*(k)\big) ,\nonumber \\
	 \mu(k) &= \frac{1}{e^{\beta\omega(k)}-1},
\end{align}
and set $(a_\beta(k))^* \equiv a_\beta^*(k)$.  This representation is due to \cite{AW}. One verifies that $[a_\beta(k), a_\beta^*(l)] = \delta(k-l)$, and that the purification of $\omega_{\r,\beta}$ is given by 
\begin{equation}
	\label{37}
	\omega_{\r,\beta}({\mathcal P}) = \scalprod{\Omega_\r}{{\mathcal P}_\beta \Omega_\r},
\end{equation}
where 
\begin{equation}
	\label{38}
	\Omega_\r =\Omega\otimes\Omega\in {\mathcal F}\otimes{\mathcal F},
\end{equation}
\blue{$\Omega$ is the vacuum vector in $\mathcal F$,} 
${\mathcal P}$ is an arbitrary polynomial in creation and annihilation operators and ${\mathcal P}_\beta$ is that same polynomial with each $a^*(k)$, $a(l)$ replaced by $a_\beta^*(k)$, $a_\beta(l)$. For the purposes of this paper, we shall call such ${\mathcal P}_\beta$  reservoir observables \footnote{In a more mathematical approach, the reservoir algebra is the Weyl algebra, represented on ${\mathcal F}\otimes{\mathcal F}$, generated by thermal Weyl operators $W_\beta(f) = \e^{\i\varphi_\beta(f)}$.}.  We denote the smoothed out operators by ($f\in L^2({\mathbb R}^3, d^3k)$) 
\begin{equation}
	\label{48}
	a_\beta^*(f) =\int_{{\mathbb R}^3} f(k) a^*_\beta(k), \quad \varphi_\beta(f) =\frac{1}{\sqrt 2} \big(a^*_\beta(f) +a_\beta(f)\big)
\end{equation}

To show that \eqref{37} is a purification of the reservoir equilibrium state, one just has to check that 
\begin{equation}
	\label{35}
	\omega_{\r,\beta} \big(a^*(k) a(l)\big) = \scalprod{\Omega_\r}{ a^*_\beta(k) a_\beta(l) \Omega_\r}
\end{equation}
equals the right side of \eqref{31}, which is easy to do. The disentangled system reservoir state is thus represented in the purification Hilbert space by the reference vector
\begin{equation}
	\label{36}
	\Psi_{\rm ref} = \Psi_\s\otimes\Omega_\r \in {\mathcal H}_{\rm ref} \equiv {\mathbb C}^N\otimes{\mathbb C}^N\otimes{\mathcal F}\otimes{\mathcal F}.
\end{equation}
The initial states we consider are exactly those which are represented by a vector (or a density matrix) on the space ${\mathcal H}_{\rm ref}$. This class contains entangled system-reservoir states. As an example, take an initial state obtained by entanglement via interaction, of the form (expressed before the continuous mode limit)  $\rho_{\s\r, 0} = e^{-\i\tau (G\otimes {\mathcal P})} ( \rho_\s\otimes\rho_{\r,\beta})  e^{\i \tau (G\otimes {\mathcal P})}$. Here, $\tau$ is a preparation time during which the disentangled $\rho_\s\otimes\rho_{\r,\beta}$ builds up entanglement due the system reservoir interaction $G\otimes {\mathcal P}$, where $G$ and ${\mathcal P}$ are self-adjoint operators (e.g. ${\mathcal P}$ a polynomial in field operators $\varphi(g)$, \eqref{5}). The purification vector of the entangled state $\rho_{\s\r,0}$ is $\Psi_{\s\r,0} = e^{-\i \tau (G\otimes\bbbone_\s\otimes{\mathcal P}_\beta)}\Psi_{\rm ref} \in {\mathcal H}_{\rm ref}$ and belongs to the class of initial states we allow.

\medskip

\blue{{\bf The glued Fock space representation.} It is sometimes useful to represent ${\mathcal F}\otimes{\mathcal F}$, where $\mathcal F$ is given in \eqref{36.f}, as a Fock space over a {\em different} single-particle space.  We explain this here and refer to \cite{MSB2}, Appendix A, for further detail and also to \cite{JP}. The symmetric Fock space ${\mathcal F}({\mathfrak H})$ over a Hilbert space $\mathfrak H$ is defined by	
$$
{\mathcal F}({\frak H}) = \oplus_{n\ge 0} \big(\otimes_{\rm sym}^n {\frak H}\big),
$$
where the summand with $n=0$ is interpreted to be $\mathbb C$ and $\otimes_{\rm sym}^n {\frak H}$ is the set of all symmetric (permutation invariant) vectors in ${\mathfrak H}\otimes\cdots\otimes {\mathfrak H}$. The exponential property of Fock spaces reads
\begin{equation}
	\label{r1}
{\mathcal F}({\mathfrak H}_1)\otimes {\mathcal F}({\mathfrak H}_2) = {\mathcal F}({\mathfrak H}_1\oplus {\mathfrak H}_2),
\end{equation}
where the equality signifies that there is an isometric isomorphism between the left and right sides. It can be easily verified using the identification
\begin{eqnarray}
	\lefteqn{a^*(f_1)\cdots a^*(f_m)\Omega\otimes a^*(g_1)\cdots a^*(g_n)\Omega }\nonumber\\
&& = a^*(f_1\oplus 0)\cdots a^*(f_m\oplus 0) a^*(0\oplus g_1)\cdots a^*(0\oplus g_n)\Omega,
\end{eqnarray}
where $f_1,\ldots f_m\in{\mathfrak H}_1$ and $g_1,\ldots f_n\in{\mathfrak H}_2$ and the vectors $\Omega$ are the vacua corresponding to the Fock spaces in question. 

With \eqref{r1} and, as per definition \eqref{36.f},  ${\mathcal F}= {\mathcal F}(L^2({\mathbb R, d^3k}))$, we have 
\begin{equation}
	\label{r2}
{\mathcal F}\otimes {\mathcal F} = {\mathcal F}\big( L^2({\mathbb R, d^3k})\oplus L^2({\mathbb R, d^3k})\big).
\end{equation}
Next, we have an identification (isometric isomorphism)  $L^2({\mathbb R, d^3k})\oplus L^2({\mathbb R, d^3k}) = L^2({\mathbb R}\times S^2, du\times d\Sigma)$, given explicitly by
\begin{equation}
\label{r3}
f\oplus g = h,\quad h(u,\Sigma) =  u
\left\{
\begin{array}{ll}
f(u,\Sigma) & u\ge 0\\
-\e^{\i \alpha} g(-u,\Sigma) & u<0
\end{array}
\right.
\end{equation}
where $\alpha\in\mathbb R$ is arbitrary (but fixed). On the right side of \eqref{r3}, the functions $f,g$ are represented in polar coordinates ${\mathbb R}^3\ni k \leftrightarrow (u,\Sigma)\in{\mathbb R}_+\times S^2$. Using the isomorphisms \eqref{r3} and \eqref{r2} we arrive at 
\begin{equation}
	\label{2.9}
{\mathcal F}\otimes {\mathcal F}= {\mathcal F}\big( L^2({\mathbb R}\times S^2, du\times d\Sigma)\big).
\end{equation}
We call the Fock space on the right side the glued Fock space, since two radial variables in ${\mathbb R}_+$ have been glued together at the origin to give a new variable $u\in\mathbb R$. Accordingly, the reference Hilbert space \eqref{36} is identified with 
\begin{equation}
{\mathcal H}_{\rm ref} = {\mathbb C}^N\otimes {\mathbb C}^N\otimes  {\mathcal F}\big( L^2({\mathbb R}\times S^2, du\times d\Sigma)\big).
\label{36.1}
\end{equation}

In the glued Fock space, the field operator $\varphi_\beta(g)$, \eqref{48}, takes the form $\varphi(g_\beta)$, where $g_\beta\in L^2({\mathbb R}\times S^2, du\times d\Sigma)$ is defined in \eqref{m1.1}. More precisely, $\varphi(g_\beta) = \frac{1}{\sqrt 2} (a^*(g_\beta) + a(g_\beta))$, where the operators on the right side are the creation and annihilation operators acting on the glued Fock space \eqref{2.9}. So, for example, $a^*(g_\beta)a^*(h_\beta)\Omega =\frac12( g_\beta\otimes h_\beta + h_\beta\otimes g_\beta)$ is a symmetric two particle state with single particle wave functions $g_\beta$ and $h_\beta$ (each a member of the enlarged single-particle Hilbert space $L^2({\mathbb R}\times S^2, du\times d\Sigma)$). 
}

\subsection{Equilibrium states} 
The uncoupled equilibrium state obtained as the continuous mode limit of $\propto e^{-\beta H_\s}\otimes e^{-\beta H_\r}$  has the  purification 
\begin{equation}
	\label{40}
	\Omega_{\s\r,\beta,0} =\Omega_{\s,\beta}\otimes\Omega_\r,
\end{equation} 
where $\Omega_\r$ is given in \eqref{38} and (see \eqref{2})
\begin{equation}
	\label{39}
	\Omega_{\s,\beta} = Z^{-1/2}_{\s,\beta} \sum_j e^{-\beta E_j/2} \phi_j\otimes\phi_j \ \ \in {\mathbb C}^N\otimes {\mathbb C}^N,
\end{equation}
with $Z_{\s,\beta} = {\rm tr} e^{-\beta H_\s}$. Of course $\Omega_{\s\r,\beta,0}\in{\mathcal H}_{\rm ref}$. The interacting equilibrium state $\Omega_{\s\r,\beta,\lambda}$, defined as the continuous mode limit of the density matrix $\propto e^{-\beta H}$ (the interacting $H$, \eqref{1}) is given by\footnote{\blue{Formula \eqref{41} is known from the perturbation theory of KMS states, see for instance \cite{BR,A,DJP, BFS}. For finite dimensional systems in particular, this is easy to understand. Let $\omega_0$ and $\omega$ be the unperturbed and perturbed equilibrium states given by density matrices $\propto\e^{-\beta H}$ and $\propto \e^{-\beta (H+V)}$. Then
\begin{eqnarray}
\omega(A)= \frac{{\rm tr}(\e^{-\beta (H+V)}A )}{{\rm tr} \e^{-\beta (H+V)}}  &=& \frac{{\rm tr}( \e^{-\beta H}  \e^{\frac\beta2 H} \e^{-\frac\beta2(H+V)}  A e^{-\frac\beta2(H+V)} \e^{\frac\beta2 H} ) }{{\rm tr} \e^{-\beta (H+V)}} \nonumber\\
&=& \frac{{\rm tr} \e^{-\beta H}}{{\rm tr} \e^{-\beta (H+V)}} \ \omega_0\big(  \e^{\frac\beta2 H} \e^{-\frac\beta2(H+V)}  A e^{-\frac\beta2(H+V)} \e^{\frac\beta2 H} \big).
\label{r7}
\end{eqnarray}
Now $\omega_0(B)=\langle \Omega_0, (B\otimes\bbbone)\Omega_0\rangle$, with $\Omega_0$ satisfying $L\Omega_0=0$, where $L=H\otimes\bbbone -\bbbone\otimes H$. We have
\begin{equation}
\big( \e^{-\frac\beta2(H+V)}  A e^{-\frac\beta2(H+V)} \big)\otimes\bbbone 
= (\bbbone\otimes\e^{-\frac\beta2 H})\big( \e^{-\frac\beta2 (L+V\otimes\bbbone)}(A\otimes\bbbone)\e^{-\frac\beta2(L+V\otimes\bbbone)}\big) (\bbbone\otimes\e^{-\frac\beta2 H}). 
\label{r8}
\end{equation}
This is so since $\e^{-\frac\beta2(L+V\otimes\bbbone)} = \e^{-\frac\beta2(H+V)}\otimes\e^{\frac\beta2H}$. It follows from \eqref{r8} that 
\begin{equation}
	\label{r9}
\omega_0\big(  \e^{\frac\beta2 H} \e^{-\frac\beta2(H+V)}  A \e^{-\frac\beta2(H+V)} \e^{\frac\beta2 H} \big) \propto \langle \Omega_0, \e^{\frac\beta2L}\e^{-\frac\beta2 (L+V\otimes\bbbone)}(A\otimes\bbbone)\e^{-\frac\beta2(L+V\otimes\bbbone)} \e^{\frac\beta2 L}\Omega_0\rangle.
\end{equation}
Since $\e^{\frac\beta2 L}\Omega_0=\Omega_0$, \eqref{r9} is of the form  $\langle \Omega, (A\otimes\bbbone)\Omega\rangle$  with $\Omega\propto \e^{-\frac\beta2 (L+V\otimes\bbbone)}\Omega_0$. Combining this with \eqref{r7} gives $\omega(A)=\langle\Omega, (A\otimes\bbbone)\Omega\rangle$. This is the formula \eqref{41}.}}
\begin{equation}
	\label{41}
	\Omega_{\s\r,\beta,\lambda}  =\frac{\e^{-\frac{\beta}{2} (L_0+\lambda G\otimes\bbbone_\s\otimes\varphi_\beta(g) )} \Omega_{\s\r,\beta,0} }{\| \e^{-\frac{\beta}{2} (L_0+\lambda G\otimes\bbbone_\s\otimes\varphi_\beta(g) )} \Omega_{\s\r,\beta,0} \| }  \in {\mathcal H}_{\rm ref}. 
\end{equation}
Here, $L_0$ is the uncoupled Liouvillian, explicitly given in \eqref{44} below. 
The equilibrium state $\Omega_{\s\r,\beta,\lambda}$, for any  $\lambda\in\mathbb R$, has the important property of cyclicity and separability\blue{, a property shared by all equilibrium (KMS) states, and which is known in generality from the theory of operator algebras \cite{BR} (Volume 2, Corollary 5.3.9). Cyclicity of $\Omega_{\s\r,\beta,\lambda}$ means that {\em any} vector $\Psi\in{\mathcal H}_{\rm ref}$ can be approximated arbitrarily well by a vector of the form $B\Omega_{\s\r,\beta,\lambda}$, for some operator $B$ which is a linear combination of terms $G\otimes\bbbone_\s\otimes {\mathcal P}_\beta$, where $G$ and ${\mathcal P}_\beta$ are system and reservoir observables}.\footnote{For any $\epsilon>0$ there is a $B$ s.t. $\|\Psi -B\Omega_{\s\r,\beta,\lambda}\|<\epsilon$.}
Separability \blue{of $\Omega_{\s\r,\beta,\lambda}$} means that an arbitrary $\Psi\in{\mathcal H}_{\rm ref}$ can also be approximated arbitrarily well by a  a vector of the form $B'\Omega_{\s\r,\beta,\lambda}$, for some operator $B'$ which is a linear combination of terms $\bbbone_\s\otimes G\otimes {\mathcal P}'_\beta$, where $G$ is a system observable and ${\mathcal P}'_\beta$ is an operator acting on ${\mathcal F}\otimes{\mathcal F}$ which commutes with any reservoir observable ${\mathcal P}_\beta$. 

The cyclicity and separating properties are easily \blue{shown} for finite dimensional systems. Namely, cyclicity comes from the fact that (in finite dimensions) any equilibirum density matrix $\e^{-\beta H_\s}$ has full range (is invertible). The separating property (which is the same as cyclicity relative to the commutant) comes about by a natural isomorphism between observables and operators commuting with observables ($X\otimes\bbbone_\s \leftrightarrow \bbbone_\s\otimes X$). Explicitly, from \eqref{39} we see that for any $k,l$,
\begin{equation}
	\label{42}
	\phi_k\otimes\phi_l = \big(G_1 \otimes\bbbone_\s\big)\Omega_{\s,\beta} = \big(\bbbone_\s\otimes G_2) \Omega_{\s,\beta},
\end{equation}
for $G_1=Z_{\s,\beta}^{1/2} e^{\beta E_l/2} |\phi_k\rangle\langle \phi_l|$ and $G_2=Z_{\s,\beta}^{1/2} e^{\beta E_k/2} |\phi_l\rangle\langle \phi_k|$. Hence in \eqref{42} we can reconstruct any basis element $\phi_k\otimes\phi_l$. By linear combination, given any $\Psi\in{\mathbb C}^N\otimes{\mathbb C}^N$, we can find $G'_1$ and $G'_2$ s.t. $\Psi = (G'_1\otimes\bbbone_\s)\Omega_{\s,\beta} = (\bbbone\otimes G'_2)\Omega_{\s,\beta}$. These properties carry over to equilibrium states of infinite dimensional (continuous mode) systems, with the only difference that exact equality might not be possible, but an arbitrarily accurate approximation of $\Psi$ can be achieved \blue{\cite{BR}}.

\medskip

{\em  Dynamics of the purified state: the Liouvillian.\ }  The uncoupled dynamics is generated by the Hamiltonian $H_0=H_\s+H_\r$, \eqref{2}, \eqref{3}. Its Heisenberg form $\e^{\i t H_0} \big(G\otimes a^*(k)\big) \e^{-\i t H_0} = \e^{\i t H_\s}Ge^{-\i t H_\s}\otimes \e^{\i\omega(k) t} a^*(k)$ is implemented in the purification Hilbert space as follows. Let $\Psi_0\in {\mathcal H}_{\rm ref}$ be the vector representing the state $\omega_0$. Then
\begin{eqnarray}
\omega_0\big( \e^{\i t H_0} \big(G\otimes a^*(k)\big) \e^{-\i t H_0} \big)
  &=& \scalprod{\Psi_0}{\big(  \e^{\i t H_\s}G \e^{-\i t H_\s}\otimes\bbbone_\s \otimes \e^{\i\omega(k) t} a_\beta^*(k)  \big)\Psi_0} \nonumber\\
	&=& \scalprod{\Psi_0}{ \e^{\i t L_0} (G\otimes\bbbone_\s\otimes a_\beta^*(k) ) \e^{-\i t L_0}\Psi_0},
	\label{43}
\end{eqnarray}
where $L_0$ is called the uncoupled Liouvillian, given by 
\begin{eqnarray}
	L_0 &=& L_\r +L_\s\nonumber\\
	L_\s &=& H_\s\otimes\bbbone_\s - \bbbone_\s\otimes H_\s\nonumber\\
	L_\r &=& H_\r\otimes\bbbone_\r -\bbbone_\r\otimes H_\r.
	\label{44}
\end{eqnarray}
Relation \eqref{43} is readily verified. Note that $L_\r\Omega_\r=0$ (see \eqref{38}). Adding the term $-\bbbone_\s\otimes H_\s$ to the system Liouvillian $L_\s$ as defined in \eqref{44} is optional.\footnote{\blue{We mean that $H_\s\otimes\bbbone_\s$ and $H_\s\otimes\bbbone_\s +\bbbone_\s \otimes K$ implement the same dynamics, no matter what the operator $K$ is. This is due to the doubling of the Hilbert space: indeed, $\e^{\i t H_\s\otimes\bbbone_\s} (G\otimes\bbbone_\s) \e^{-\i H_\s\otimes\bbbone_\s}= \e^{\i t(H_\s\otimes\bbbone_\s +\bbbone_\s \otimes K) } (G\otimes\bbbone_\s) \e^{-\i ( H_\s\otimes\bbbone_\s +\bbbone_\s \otimes K)}$, because $\e^{\i t( H_\s\otimes\bbbone_\s +\bbbone_\s \otimes K)} = \e^{\i t H_\s}\otimes  \e^{\i t K}$. The observables are always of the form $G\otimes \bbbone_\s$ acting trivially on the second factor. This is why we can modify the generator by adding a term acting on the second tensor factor without changing the dynamics.
}}
It serves to ensure the agreeable property  $L_\s\Omega_{\s,\beta}=0$ (see \eqref{39}). Thus we have 
\begin{equation}
\label{m2}
L_0\Omega_{\s\r,\beta,0} =0.
\end{equation} 
The full, interacting dynamics generated by $H$, \eqref{1}, is implemented  as
\begin{equation}
	\omega_0 \big( \e^{\i t H} (X\otimes {\mathcal P}) \e^{-\i tH}\big)
=  \scalprod{\Psi_0}{\e^{\i t L_\lambda} (X\otimes \bbbone_\s\otimes {\mathcal P}_\beta) \e^{-\i t L_\lambda}\Psi_0}.
	\label{46}
\end{equation}
Here,  $L_\lambda$ is the coupled Liouvillian, given by
\begin{eqnarray}
	L_\lambda&=& L_0 +\lambda I\nonumber\\
	I &=& G\otimes\bbbone_\s\otimes\varphi_\beta(g) - J\big(G\otimes\bbbone_\s \otimes \varphi_\beta(g)\big) J.
	\label{45}
\end{eqnarray}
\blue{The operator $L_\lambda$ is self-adjoint, for any value of $\lambda\in\mathbb R$.  This is proven for instance by using {\em Glimm-Jaffe-Nelson triples} techniques, {\em c.f.} \cite{FM}. }

We will not use explicitly the form of $L_\lambda$ in this paper, but let us explain the term $J\big(G\otimes\bbbone_\s \otimes \varphi_\beta(g)\big) J$ in \eqref{45}. This is an operator which commutes with all observables (i.e., with all operators which are linear combinations of the form $X\otimes\bbbone_\s\otimes {\mathcal P}_\beta$). The map $J$ is an anti-unitary involution\blue{, the modular conjugation of Tomita Takesaki theory, defined by the property
\begin{equation}
\label{m3}
J \e^{-\beta L_0/2} \big(  G\otimes\bbbone_\s\otimes {\mathcal P}_\beta\big) \Omega_{\s\r,\beta,0} = \big(  G\otimes\bbbone_\s\otimes {\mathcal P}_\beta\big)^* \Omega_{\s\r,\beta,0},
\end{equation}
valid for all $G$ and ${\mathcal P}_\beta$.\footnote{ \blue{$J\e^{-\beta L_0/2}$ is the polar decomposition of the antilinear operator $S$ defined by $S A\Omega_{\s\r,\beta,0} = A^*\Omega_{\s\r,\beta,0}$ for all observables $A$, see for instance \cite{BR} (Volume 1, Definition 2.5.10).}}  The action of $J$ can be written down explicitly. Namely, $J=J_\s\otimes J_\r$, with $J_\s$, $J_\r$ defined by the following relations (plus antilinear extension and continuity)
\begin{eqnarray}
J_\s \big( \chi_1	\otimes\chi_2\big)  &=& {\cal C}\chi_2 \otimes {\cal C}\chi_1 \nonumber\\
J_\r \big(  \psi_1(k_1,\ldots,k_m)\otimes\psi_2(\ell_1,\ldots,\ell_n) \big) &=& \overline{\psi_2}(\ell_1,\ldots,\ell_n)\otimes \overline{\psi_1}(k_1,\ldots,k_m),
\label{m6}
\end{eqnarray}
where $\chi_1$, $\chi_2 \in {\mathbb C}^N$ and $\cal C$ is the antiunitary taking complex conjugates of coordinates in the eigenbasis of $H_\s$.  In \eqref{m6}, the $\psi_{1,2}\in{\mathcal F}$ are finite particle wave functions and $\overline{\psi_{1,2}}$ their complex conjugates. For more detail we refer {\em e.g.} to \cite{MSB2, BFS}. It follows from \eqref{m6} and \eqref{44} that 
\begin{equation}
\label{m4}
J L_0 J = -L_0.
\end{equation}
We will not use the fine properties of $J$ in this paper. A important property of $J$ that we will use is this:} given any system observable $A$ and any reservoir observable ${\mathcal P}_\beta$, the operator $J (A\otimes\bbbone_\s\otimes{\mathcal P}_\beta)J$ {\em commutes} with all system-reservoir observables $B\otimes\bbbone_\s\otimes {\mathcal Q}_\beta$. Adding the commuting term $J\big(G\otimes\bbbone_\s \otimes \varphi_\beta(g)\big) J$ in the interaction is optional (meaning that the equality \eqref{46} still holds if $I$ is defined without adding this term). The reason for this non-uniqueness of the Liouvillian comes from the fact that adding to the generator an operator which commutes with all observables will not alter the dynamics of observables. The choice \eqref{45} ensures that the coupled equilibrium state \eqref{41} satisfies 
\begin{equation}
	\label{47}
	L_\lambda \Omega_{\s\r,\beta,\lambda} =0.
\end{equation}
\blue{To prove \eqref{47}, denote the operator $\lambda I$ in \eqref{45} by $V-JVJ$, defining $V=G\otimes\bbbone_\s\otimes\varphi_\beta(g)$. Taking into account \eqref{41}, which reads $\Omega_{\s\r,\beta,\lambda} \propto \e^{-\beta(L_0+V)/2}\Omega_{\s\r,\beta,0}$, we have
\begin{eqnarray}
L_\lambda \Omega_{\s\r,\beta,\lambda} &\propto& (L_0+V-JVJ) \e^{-\beta(L_0+V)/2}\Omega_{\s\r,\beta,0} \nonumber\\
&=& \e^{-\beta(L_0+V)/2} (L_0+V) \Omega_{\s\r,\beta,0} -JVJ \e^{-\beta(L_0+V)/2} \Omega_{\s\r,\beta,0}\, .
\label{m1}
\end{eqnarray}
Due to \eqref{m2}, we have $\e^{-\beta(L_0+V)/2} \Omega_{\s\r,\beta,0}=\e^{-\beta(L_0+V)/2} \e^{\beta L_0/2} \Omega_{\s\r,\beta,0}$ and one can expand the product of the last two exponentials into an (imaginary time) Dyson series with general term $(-1)^n \int_{0\le t_n\le\cdots \le t_1\le\beta/2} V(t_n)\cdots V(t_1) dt_1\cdots dt_n$, where $V(t) = \e^{-t L_0} Ve^{t L_0}$. As mentioned above, $JVJ$ commutes with $V(t)$ and hence
\begin{equation}
\label{m5}
 JVJ  \e^{-\beta(L_0+V)/2} \Omega_{\s\r,\beta,0} = \e^{-\beta(L_0+V)/2}  e^{\beta L_0/2} JVJ \Omega_{\s\r,\beta,0} = \e^{-\beta(L_0+V)/2}  V \Omega_{\s\r,\beta,0}.
\end{equation}
The last equality is true since $J\Omega_{\s\r,\beta,0} = \Omega_{\s\r,\beta,0}$ (see \eqref{m3}) and $\e^{\beta L_0/2} J =  J \e^{-\beta L_0/2}$ (see \eqref{m4}) and since $J\e^{-\beta L_0/2} V\Omega_{\s\r,\beta,0} = V\Omega_{\s\r,\beta,0}$, by \eqref{m3} again and since $V$ is self-adjoint. Using \eqref{m5} in \eqref{m1} (and $L_0\Omega_{\s\r,\beta,0}=0$) shows that the right hand side of \eqref{m1} vanishes. Hence \eqref{47} is proven. 
}

\subsection{Representation of the dynamics}
\label{secrepdyn} The Heisenberg evolution of a system observable $X$ is 
\begin{equation}
	\label{29}
	\alpha^t_\lambda(X\otimes\bbbone_\r) =e^{\i tH}( X\otimes\bbbone_\r)e^{-\i tH},
\end{equation}
where $H$ is the interacting system-reservoir Hamiltonian \eqref{1}. Let $\omega_0$ be an (initial) system-reservoir state, with purification  $\Psi_0\in{\mathcal H}_{\rm ref}$.  The vector $\Psi_0$ can be approximated arbitrarily well by $B'\Omega_{\s\r,\beta,\lambda}$ for a suitable $B'$ commuting with all observables. \blue{This follows from the separability property of the state $\Omega_{\s\r,\beta,\lambda}$, as explained before \eqref{42}.} Since the full dynamics is unitary, this approximation is uniform in time. We will hence assume without loss of generality that
\begin{equation}
	\label{56}
	\Psi_0 = B'\Omega_{\s\r,\beta,\lambda}.
\end{equation} 
Note that if the initial state is of the form $\rho_\s\otimes\omega_{\r,\beta}$ then  \blue{the corresponding vector is $\Psi_0 = \Omega_\s\otimes\Omega_\r$ for some $\Omega_\s\in {\mathcal H}_\s\otimes{\mathcal H}_\s$ and where $\Omega_\r$ is given in \eqref{38}. Then there is an operator $B'_\s\in{\mathcal B}({\mathcal H}_\s)$ such that $\Omega_\s = (\bbbone_\s\otimes B'_\s) \Omega_{\s,\beta}$, see the discussion involving \eqref{42}. Furthermore,  by \eqref{41}, $\Omega_{\s\r,\beta,\lambda}= \Omega_{\s,\beta}\otimes\Omega_{\r}+O(\lambda)$, so we have \eqref{56} with} 
\begin{equation}
	\label{56.1}
	B'=\bbbone_\s\otimes B'_\s\otimes \bbbone_\r +O(\lambda),\qquad \mbox{some $B'_\s\in{\mathcal B}({\mathcal H}_\s)$}.
\end{equation}
\blue{Here, ${\mathcal B}({\mathcal H}_\s)$ denotes the set of all bounded operators on ${\mathcal H}_\s$.} What follows works for all initial states \eqref{56}. We have
\begin{eqnarray}
\omega_0\big(\alpha^t_\lambda (X\otimes\bbbone_\r)\big)
 &=& \scalprod{\Psi_0}{e^{\i t L_\lambda} (X\otimes\bbbone_\s\otimes\bbbone_\r)e^{-\i t L_\lambda}\Psi_0} \nonumber\\
	&=& \scalprod{\Psi_0}{B'e^{\i t L_\lambda}  (X\otimes\bbbone_\s\otimes\bbbone_\r)e^{-\i t L_\lambda}\Omega_{\s\r,\beta,\lambda}} \nonumber\\
	&=&\scalprod{\Psi_0}{B' e^{\i t L_\lambda}  (X\otimes\bbbone_\s\otimes\bbbone_\r) \Omega_{\s\r,\beta,\lambda}}.
	\label{33}
\end{eqnarray}
In the second equality we moved $B'$ to the left, as it commutes with the observable $\e^{\i t L_\lambda} (X\otimes\bbbone_\s\otimes\bbbone_\r)\e^{-\i t L_\lambda}$. 
In the third we use the invariance \eqref{47}. Next comes the core analytical tool, the resonance expansion of $e^{\i t L_\lambda}$. It is important to realize that this expansion is only correct in the weak sense; one cannot perform it independently on both factors $e^{\pm \i tL_\lambda}$ in \eqref{33}.\footnote{This is readily seen: weakly, $\e^{\i tL_\lambda}\rightarrow |\Omega_{\s\r,\beta,\lambda}\rangle\langle \Omega_{\s\r,\beta,\lambda}|$ for $t\rightarrow\infty$ and using this for both propagators in \eqref{33} would yield the result  $\scalprod{ \Omega_{\s\r,\beta,\lambda}}{(X\otimes\bbbone_\s\otimes\bbbone_\r)\Omega_{\s\r,\beta,\lambda}}  |\langle \Omega_{\s\r,\beta,\lambda},\Psi_0\rangle|^2$ for $t\rightarrow\infty$. But this is not the correct final state.} This is why we have to exploit the algebraic structure (existence of $B'$) and eliminate one of the propagators $e^{-\i tL_\lambda}$ by making it act on the invariant state $\Omega_{\s\r,\beta,\lambda}$ in \eqref{33}.

The right side of \eqref{33} is of the form $\scalprod{\psi}{e^{\i t L_\lambda}\phi}$ for two vectors $\psi$, $\phi$. We use the usual resolvent  representation of the propagator,
\begin{equation}
	\label{49}
	\scalprod{\psi}{\e^{\i t L_\lambda}\phi} = \frac{-1}{2\pi \i}\int_{{\mathbb R}-\i} \e^{\i t z} \scalprod{\psi}{(L_\lambda-z)^{-1}\phi} dz.
\end{equation}
The integral is over the horizontal contour $z=x-\i$, $x\in{\mathbb R}$. Since $L_\lambda$ is self-adjoint, $(L_\lambda-z)^{-1}$ is a well defined, bounded operator. We explain the further analysis of \eqref{49} in the technically easiest situation (which requires the most regularity, though), namely, when the {\em spectral deformation} technique applies. The strategy  is to construct a meromorphic continuation in $z$ of the function $\scalprod{\psi}{(L_\lambda-z)^{-1}\phi}$, extending the domain of $z$ from the lower half plane ${\mathbb C}_-$ across the real axis into (parts of) the upper complex half plane. Whether this is possible depends of course on the operator $L_\lambda$ (and the vectors $\psi,\phi$). 
\medskip

Denote by $U_\theta$ the action of $T_\theta$ defined in condition (A), lifted from the single-particle space to Fock space. Then $U_\theta$, $\theta\in\mathbb R$, is a unitary group on ${\mathcal H}_{\rm ref}$ \eqref{36} (or equivalently, by isometric isomorphy, \eqref{36.1}) satisfying  
\begin{equation}
	\scalprod{\psi}{(L_\lambda-z)^{-1}\phi} = \scalprod{U_\theta \psi}{U_\theta (L_\lambda-z)^{-1}\phi}
	=  \scalprod{\psi_{\bar \theta}}{(L_{\lambda,\theta}-z)^{-1}\phi_\theta}
	\label{50}
\end{equation}
and (assuming condition (A) above), the right side of  \eqref{50} has an extension to complex values of $\theta$ (here, $\bar\theta$ is the complex conjugate of $\theta$ and it shows up in \eqref{50} since the scalar product is antilinear in its left argument).  The first equality in \eqref{50} is due to unitarity of $U_\theta$ and we define $\psi_\theta=U_\theta \psi$, $\phi_\theta=U_\theta \phi$ and $L_{\lambda,\theta} =U_\theta L_\lambda U^*_\theta$. \blue{The deformed Liouvillian $L_{\lambda,\theta}$ is of the form $L_{\lambda,\theta} = L_{0,\theta} +\lambda I_\theta$, acting on ${\mathcal H}_{\rm ref}$ \eqref{36.1}, where 
\begin{equation}
	\label{r4}
U_\theta L_0 (U_\theta)^*	= L_0 +\theta N,
\end{equation}
and $N$ is the number operator on the glued Fock space \eqref{2.9}, that is, $N=d\Gamma(\bbbone)$. Equality \eqref{r4} follows from the explicit identifications given above following \eqref{r1}. 
}

The relation \eqref{50} stays valid for complex values of $\theta$ due to the identity theorem of complex analysis (varying the real part of $\theta$ does not change the inner products, due to unitarity). When $\theta$ becomes complex, $L_{\lambda,\theta}$ is {\em not} a self-adjoint operator any longer (it is not even a normal operator) and hence generically, its spectrum leaves the real axis as ${\rm Im}\theta\neq 0$. Take now $\theta$ with ${\rm Im}\theta=\theta_0>0$ fixed.
\begin{center}
	\label{figure1}
\includegraphics[width=15cm]{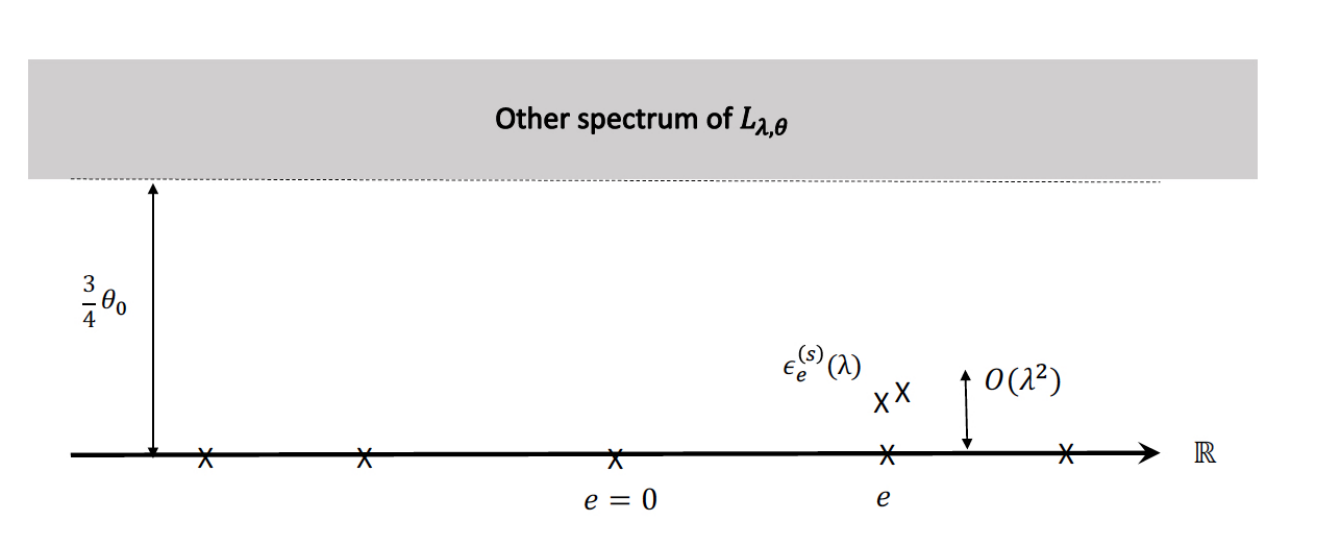}	
\end{center}
\centerline{\small Fig.1: The eigenvalues $e$ of $L_{0,\theta}$ bifurcate into eigenvalues $\epsilon_e^{(s)}(\lambda)$  of $L_{\lambda,\theta}$ for nonzero $\lambda$.}
\bigskip
\bigskip

\noindent
By analytic perturbation theory and the fact that $L_{0,\theta} = L_0 +\theta N$, where $N$ is the number operator, having spectrum ${\mathbb N}\cup\{0\}$, one shows the following result \cite{JP,BFS,MSB2}: 

\smallskip
{\em  In a strip $\{ z\in{\mathbb C}\ :\ 0\le {\rm Im}z < \theta_0/2\}$, the spectrum of the operator $L_{\lambda,\theta} = L_{0,\theta} +\lambda I_\theta$ (c.f. \eqref{45}) consists of eigenvalues which are independent of $\theta$ (for $\lambda$ not too large compared to $\theta$).  All other spectrum of $L_{\lambda,\theta}$ is located within $\{z\in{\mathbb C}\ :\ {\rm Im}z >3\theta_0/4\}$.}

\smallskip
The situation is depicted in Fig.1.  For $\lambda=0$, the eigenvalues coincide (including multiplicity) with those of $L_\s$. More precisely, the (rank $m_e$) spectral projection associated to the eigenvalue $e$ of $L_{0,\theta}$ is given by
\begin{equation}
	\label{57}
	P_e = P(L_\s=e)\otimes P_\r,
\end{equation}
where $P(L_\s=e)$ is the eigenprojection of $L_\s$ associated to the eigenvalue $e$ and $P_\r = |\Omega_\r\rangle\langle \Omega_\r|$. Since $e$ is an isolated eigenvalue of $L_{0,\theta}$, analytic perturbation theory implies that  for small $\lambda$, $e$ splits into $\le m_e$ eigenvalues $\epsilon_e^{(s)}(\lambda)$, $s=1,2,\ldots$ (the added up multiplicity equaling $m_e$), which are analytic at $\lambda=0$ and have the expansion
\begin{equation}
	\label{64}
	\epsilon_e^{(s)}(\lambda) = e +\lambda^2 a_e^{(s)} +O(\lambda^4).
\end{equation}
The corrections $a_e^{(s)}$ can be calculated by perturbation theory \footnote{ In principle, there are $O(\lambda)$ correction terms given by $P_e I_\theta P_e$, but this operator vanishes for the interactions we consider.}. They are the eigenvalues of the {\em level shift operator} 
\begin{equation}
	\label{59}
	\Lambda_e = - P_e I P_e^\perp (L_0 -e +\i 0)^{-1} I P_e.
\end{equation}
\blue{This is a fact from second order analytic perturbation theory, sometimes also phrased as the Feshbach map, see for instance \cite{Mlso, BFS,KoMeDyn}.} 
Using \eqref{50} in \eqref{49} yields
\begin{equation}
\scalprod{\psi}{\e^{\i t L}\phi} =	 \frac{-1}{2\pi \i}\sum_{e\in{\rm spec}(L_\s)} \sum_{s=1}^{m_e} \oint_{\Gamma_e^{(s)}} \e^{\i t z} \scalprod{\psi_{\bar \theta}}{(L_{\lambda, \theta}-z)^{-1}\phi_\theta} dz +O\big( \e^{- \frac 34\theta_0 t}\big).
\label{51}
\end{equation}
To arrive at \eqref{51}, we have deformed the contour of integration $z=x-\i$ into $z=x+\tfrac34\i\theta_0$, thereby (by the residue theorem) creating the contour integrals $\oint_{\Gamma_e^{(s)}}$, where $\Gamma_e^{(s)}$ is a circle centered at $\epsilon_e^{(s)}(\lambda)$, not containing any other eigenvalue of $L_{\lambda,\theta}$. The remainder decays at rate $-\tfrac34 \theta_0$ due to the factor $\e^{\i t z}$. \blue{Indeed, this remainder is a contour integral over $z=x+\frac34\i\theta_0$ and for such $z$, we have $|\e^{\i t z}| = \e^{-\frac34\theta_0 t}$. }  Consider the situation where all of the $\epsilon_e^{(s)}$ are distinct (for $\lambda\neq 0$). The integrand in \eqref{51} has a simple pole at $z=\epsilon_e^{(s)}$ in the interior of $\Gamma_e^{(s)}$ and so we have 
\begin{equation}
\frac{-1}{2\pi\i}\oint_{\Gamma_e^{(s)}} \e^{\i t z}(L_{\lambda,\theta}-z)^{-1} dz=  \e^{\i t \epsilon_e^{(s)}(\lambda) }\big( \frac{-1 }{2\pi\i}\big) \oint_{\Gamma_e^{(s)}} (L_{\lambda,\theta}-z)^{-1} dz
\equiv \e^{\i t \epsilon_e^{(s)}(\lambda)} \Pi_e^{(s)},
	\label{65}
\end{equation}
where $\Pi_e^{(s)} = \Pi_e^{(s)}(\lambda,\theta)$ is the (Riesz) spectral  projection associated to the eigenvalue $\epsilon_e^{(s)}(\lambda)$ of $L_{\lambda,\theta}$. 
Combining \eqref{33}, \eqref{51}
and \eqref{65} yields
\begin{eqnarray}
\omega_0\big(\alpha_\lambda^t(X\otimes{\bbbone}_\r)\big )  &=& \sum_{e\in{\rm spec}(L_\s)} \sum_{s=1}^{m_e} \e^{\i t \epsilon_e^{(s)}}  \scalprod{[(B')^*\Psi_0]_{\bar\theta}}{\Pi_e^{(s)} \big( X\otimes{\bbbone}_\s\otimes {\bbbone}_\r \big) [\Omega_{\s\r,\beta,\lambda}]_\theta}\nonumber\\
&&+O\big( \lambda\e^{- \frac 34\theta_0 t}\big).
		\label{54}
\end{eqnarray}
Note that the remainder\blue{, which is given by
\begin{equation}
	\label{r6}
\int_{\mathbb R} \e^{\i(x+3\i\theta_0/4)}  \langle [(B')^*\Psi_0]_{\bar\theta},  (L_{\lambda,\theta} - x-3\i\theta_0/4)^{-1} \big( X\otimes{\bbbone}_\s\otimes {\bbbone}_\r \big) [\Omega_{\s\r,\beta,\lambda}]_\theta\rangle dx,
\end{equation}
vanishes to zeroth order in $\lambda$. 	This is so since to this order, $[\Omega_{\s\r,\beta,\lambda}]_\theta$ is given by $\Omega_{\s,\beta}\otimes\Omega_\r$ and the $L_{\lambda,\theta}$ in the resolvent is simply $L_\s$, to this order and when applied to the vector in question. The contour integral \eqref{r6} is thus not enclosing any singularities of the integrand to order $\lambda^0$ and so it vanishes.} If the initial state is of the form $\rho_\s\otimes\omega_{\s,\beta}$, then the remainder in \eqref{54} is actually $O(\lambda^2)$, due to \eqref{56.1} (see Proposition 4.2 of \cite{MSB1}).

Our next step is to eliminate the $\theta$ dependence of the main term in \eqref{54}. Consider first $e=0$. Due to \eqref{47} and since $[\Omega_{\s\r,\beta,\lambda}]_\theta = U_\theta \Omega_{\s\r,\beta,\lambda}$ is analytic in $\theta$, we have $L_{\lambda,\theta}[\Omega_{\s\r,\beta,\lambda}]_\theta=0$. It follows that  $L_{\lambda,\theta}$ has an eigenvalue $\epsilon_0^{(1)}=0$ for all $\lambda,\theta$. We use $s=1$ to label it. The associated eigenprojection is
\begin{equation}
	\label{53}
	\Pi_0^{(1)}  = |[\Omega_{\s\r,\beta,\lambda}]_\theta\rangle \langle [\Omega_{\s\r,\beta,\lambda}]_{\bar\theta}|. 
\end{equation}
In the sum \eqref{54}, the term $e=0$, $s=1$ equals
\begin{eqnarray}
	\lefteqn{
		\scalprod{[(B')^*\Psi_0]_\theta}{[\Omega_{\s\r,\beta,\lambda}]_{\theta}} \scalprod{[\Omega_{\s\r,\beta,\lambda}]_{\bar\theta}}{\big(X\otimes{\bbbone}_\s\otimes {\bbbone}_\r\big)[\Omega_{\s\r,\beta,\lambda}]_\theta} }\nonumber\\
	&=& \scalprod{\Psi_0}{B'\Omega_{\s\r,\beta,\lambda}} \scalprod{\Omega_{\s\r,\beta,\lambda}}{\big(X\otimes{\bbbone}_\s\otimes {\bbbone}_\r\big) \Omega_{\s\r,\beta,\lambda}}\nonumber\\
	&=&{\rm tr}_\s \big(\rho_{\s,\beta,\lambda} X\big).
	\label{55}
\end{eqnarray}
The first equality in \eqref{55} holds by the identity principle of complex analysis. The final equality follows from (recall \eqref{56}) $\scalprod{\Psi_0}{B'\Omega_{\s\r,\beta,\lambda}} = \scalprod{\Psi_0}{\Psi_0}=1$ and from the definition of $\rho_{\s,\beta,\lambda}$ as  the reduction to the system of the full, interacting system-reservoir equilibrium state. Above, we are able to arrive at the result \eqref{55}, which is non-perturbative in $\lambda$, since we know to begin with that $L_\lambda \Omega_{\s\r,\beta,\lambda} =0$.

For the other terms in the sum \eqref{54}, associated with nonzero resonance energies, we use regular analytic perturbation theory in $\lambda$ (as we do not know an {\em a priori} expression for them). Consider the situation where each $\Lambda_e$ is diagonalizable, i.e.,
\begin{equation}
	\label{61}
	\Lambda_e = \sum_{s=1}^{m_e} a_e^{(s)}Q_e^{(s)},
\end{equation}
where $a_e^{(s)}$ and $Q_e^{(s)}$ are the eigenvalues and rank-one eigenprojections, neither depending on $\theta$. We have 
\begin{equation}
	\label{62}
	Q_e^{(s)} < P(L_\s=e)\qquad \mbox{and}\qquad \sum_{s=1}^{m_e} Q_e^{(s)} = P(L_\s =e).
\end{equation}
The relation $L_\lambda \Omega_{\s\r,\beta,\lambda}=0$ implies that $\Lambda_0\Omega_{\s,\beta}=0$. \blue{This follows from the isospectrality property of the Feshbach map, see {\em e.g.} Theorem B.1 in \cite{KoMeDyn}.}  Assuming that all the eigenvalues of $\Lambda_0$ are simple then yields
\begin{equation}
	\label{125}
	Q_0^{(1)} = |\Omega_{\s,\beta}\rangle\langle \Omega_{\s,\beta}|.
\end{equation}
Analytic perturbation theory gives the following expansion for $\Pi_e^{(s)}$, the spectral projection of $L_{\lambda,\theta}$ associated to $\epsilon_e^{(s)}$ 
\begin{equation}
	\label{63} 
	\Pi_e^{(s)}(\theta, \lambda) = Q_e^{(s)} \otimes |\Omega_\r\rangle\langle\Omega_\r| +O(\lambda).
\end{equation}
Consider a term in \eqref{54} with $(e,s)$ fixed (not equal to $(0,1)$). We have
\begin{eqnarray}
	\lefteqn{
		\scalprod{[(B')^*\Psi_0]_{\bar\theta}}{\Pi_e^{(s)} \big( X\otimes{\bbbone}_\s\otimes {\bbbone}_\r \big) [\Omega_{\s\r,\beta,\lambda}]_\theta}}\nonumber\\
	&=& \scalprod{\Psi_0}{B' \big(Q_e^{(s)}\otimes|\Omega_\r\rangle \langle\Omega_\r|  \big) \big( X\otimes{\bbbone}_\s\otimes {\bbbone}_\r \big) \Omega_{\s\r,\beta,\lambda}}+O(\lambda)\nonumber\\
	&=& \scalprod{\Psi_0}{B' \big(Q_e^{(s)}\otimes\bbbone_\r \big) \big( X\otimes{\bbbone}_\s\otimes {\bbbone}_\r \big) (\Omega_{\s,\beta}\otimes \Omega_\r)}+O(\lambda). 
	\label{66}
\end{eqnarray}
In the first equality of \eqref{66} we have used the approximation \eqref{63} and that $U_\theta \Omega_\r=\Omega_\r$. In the second equality we made use of $(\bbbone_\s\otimes\bbbone_\s\otimes |\Omega_\r\rangle\langle \Omega_\r|) \Omega_{\s\r,\beta,\lambda} = \Omega_{\s,\beta}\otimes\Omega_\r +O(\lambda^2)$  (see \eqref{40} and  \eqref{41}). If the initial condition is of the form $\rho_\s\otimes\omega_{\r,\beta}$, then \eqref{56.1} holds and it is not hard to see that since $\langle \Omega_{\r}|  I |\Omega_\r\rangle=0$,  the remainder in \eqref{66} is actually $O(\lambda^2)$.  Due to the cyclicity of  $\Omega_{\s,\beta}$, there are uniquely defined operators ${\mathcal Q}_e^{(s)}$ acting on system observables, satisfying 
\begin{equation}
	\label{78}
	\big(  {\mathcal Q}_e^{(s)}(X)\otimes\bbbone_\s \big)\Omega_{\s,\beta} = Q_e^{(s)}(X\otimes\bbbone_\s)\Omega_{\s,\beta},\quad \forall X.
\end{equation}
The $ {\mathcal Q}_e^{(s)}$ are a family of disjoint projection operators (as the $Q_e^{(s)}$ are). The main term on the right side of \eqref{66} is then
\begin{eqnarray}
	\lefteqn{
\scalprod{\Psi_0}{B' \big(Q_e^{(s)}\otimes\bbbone_\r \big) \big( X\otimes{\bbbone}_\s\otimes {\bbbone}_\r \big) (\Omega_{\s,\beta}\otimes \Omega_\r)}}\nonumber\\
	&=&
	\scalprod{\Psi_0}{ \big( {\mathcal Q}_e^{(s)}(X)\otimes\bbbone_\s\otimes\bbbone_\r \big) B'  (\Omega_{\s,\beta}\otimes \Omega_\r)}\nonumber\\
	&=&
	\omega_0\big( {\mathcal Q}_e^{(s)}(X)\otimes\bbbone_\r\big)
	+O(\lambda)
	\label{82}
\end{eqnarray}
To arrive at \eqref{82}, we have used that $B'$ commutes with all observables, so we can move it to the right of ${\mathcal Q}_e^{(s)}(X)\otimes\bbbone_\s\otimes\bbbone_\r$ and we also take into account that 
\begin{equation}
	\label{71}
	B'(\Omega_{\s,\beta}\otimes\Omega_\r) = B' \Omega_{\s\r,\beta,\lambda} +O(\lambda) = \Psi_0 +O(\lambda).
\end{equation}
The $O(\lambda)$ term in \eqref{71} comes about by replacing the uncoupled equilibrium $\Omega_{\s,\beta}\otimes\Omega_\r$ by the coupled one, $\Omega_{\s\r,\beta,\lambda}$. The initial state $\Psi_0$ emerges in \eqref{71} due to \eqref{56}. Again, for initial states $\rho_\s\otimes\omega_{\r,\beta}$, the remainder in \eqref{82}, \eqref{71} is actually $O(\lambda^2)$, due to \eqref{56.1}. 

Combining \eqref{82} with \eqref{66}, \eqref{55} and \eqref{54} shows the expansion
\begin{eqnarray}
\omega_0\big(\alpha_\lambda^t(X\otimes{\bbbone}_\r)\big ) &=& {\rm tr}_\s\big(\rho_{\s,\beta,\lambda} X\big) +  \sum_{(e,s)\neq (0,1)} \e^{\i t \epsilon_e^{(s)}}  \omega_0\big({\mathcal Q}_e^{(s)}(X)\otimes\bbbone_\r \big)\nonumber\\
	&& + \ O(\lambda \e^{-\gamma(\lambda)t})+ O\big( \lambda \e^{- \frac 34\theta_0 t}\big).
	\label{83}
\end{eqnarray}
\blue{Here,  $\gamma(\lambda)$ was defined in \eqref{77} to be the slowest decay rate.  The corresponding error term in \eqref{83} stems from making in \eqref{54} approximations to within $O(\lambda)$ in the scalar product, which is time independent. Since $\gamma(0)=0$ and $\lambda\mapsto\gamma(\lambda)$ is continuous, we have $\gamma(\lambda) < \tfrac34\theta_0$ for small enough $\lambda$ and so} the second error term in \eqref{83} is smaller than the first one. Equation \eqref{83} is the basic result of the resonance theory for system observables. Again, as explained during the derivation, for initial states $\rho_\s\otimes\omega_{\r,\beta}$ the $\lambda$ in both remainders in \eqref{83} are actually $\lambda^2$.

\subsection{Proof of \eqref{14}}
Suppose that the initial state is disentangled, $\omega_0 = \omega_\s\otimes\omega_{\r,\beta}$, where $\omega_\s$ is given by a general system density matrix $\rho$ and $\omega_{\r,\beta}$ is the reservoir equilibrium (or a local perturbation thereof). The remainders in \eqref{83} are then $O(\lambda^2)$. The dynamical map $\rho\mapsto V_t\rho$ is defined by
\begin{equation}
	\label{73}
	{\rm tr}_\s\big( (V_t\rho) \, X\big) = \omega_0\big( \alpha_\lambda^t(X\otimes\bbbone_\r)\big),\qquad \forall X.
\end{equation}
The result \eqref{83} then implies
\begin{equation}
	\label{74}
	V_t\rho = \rho_{\s,\beta,\lambda} +W_t\rho +O(\lambda^2 \e^{-\gamma(\lambda) t}),
\end{equation}
where $W_t$ is the map on density matrices defined by duality. It is given by \eqref{15} in which the sum is over $j=(e,s)\neq (0,1)$. In particular, the ${\mathcal P}_e^{(s)}$ are determined uniquely by 
\begin{equation}
	\label{79}
	{\rm tr} ({\mathcal P}_e^{(s)}\rho) X = {\rm tr} \rho ({\mathcal Q}_e^{(s)} X),\qquad \forall\rho, X.
\end{equation}
Recall the definition \eqref{78}, in which the $Q_e^{(s)}$ are spectral projections of the level shift operators \eqref{61}. They form a family of disjoint projections, $Q_e^{(s)} Q_{e'}^{(s')} = \delta_{e,e'} \delta_{s,s'} Q_e^{(s)}$ and satisfy (see \eqref{62}) $\sum_{(e,s)\neq (0,1)} Q_e^{(s)} = \bbbone_\s\otimes\bbbone_\s - |\Omega_{\s,\beta}\rangle \langle \Omega_{\s,\beta}|$. Accordingly, it follows from \eqref{78} that 
\begin{equation}
	\label{79.1}
	{\mathcal Q}_e^{(s)} {\mathcal Q}_{e'}^{(s')} = \delta_{e,e'} \delta_{s,s'} {\mathcal Q}_e^{(s)}
\end{equation}
and $\forall X\in {\mathcal B}({\mathcal H}_\s)$,
\begin{equation}
	 \sum_{(e,s)\neq (0,1)} {\mathcal Q}_e^{(s)} X= X- {\rm tr}(\rho_{\s,\beta,0}\,  X)\bbbone_\s. 
	\label{79.2}
\end{equation} 
The duality \eqref{79} then translates into the corresponding properties \eqref{15.1} of the family ${\mathcal P}_e^{(s)}$. 

\section{Derivation of the main results}

\subsection{Proof of \eqref{24}} 
\label{secother}Define the operator ${\mathcal M}(\lambda)$, acting on system observables, by its spectral decomposition
\begin{equation}
	\label{67}
	{\mathcal M}(\lambda) = \sum_{(e,s)\neq (0,1) } \  \epsilon_e^{(s)}(\lambda)\ {\mathcal  Q}_e^{(s)},
\end{equation}
where the sum is over all $e,s$ except $(e,s)=(0,1)$. Note that if $\epsilon_e^{(s)}(\lambda)\neq 0$ for $(e,s)\neq(0,1)$ (this is typically the case and holds in particular if the Fermi Golden Rule \eqref{22} is satisfied), then we have \footnote{Note that $Q_0^{(1)} (X\otimes\bbbone_\s)\Omega_{\s,\beta} = ({\rm tr}\rho_{\s,\beta,0}X) \Omega_{\s,\beta}$, so by \eqref{78} ${\mathcal Q}_0^{(1)}(X) = {\rm tr}(\rho_{\s,\beta,0} X)\bbbone_\s$.} 
\begin{equation}
	\label{85}
	{\rm ker } {\mathcal M}(\lambda) = {\rm ran} {\mathcal Q}_0^{(1)}
	= \{ {\mathbb C} \rho_{\s,\beta,0} \}^\perp
	\equiv \{ X\, :\, {\rm tr}(\rho_{\s,\beta,0} X)=0\}.  
\end{equation}
Using the definition \eqref{67}, the power series expansion of the exponential and \eqref{79.2}, we obtain
\begin{eqnarray}
	\e^{\i t {\mathcal M}(\lambda)} &=& \sum_{(e,s)\neq (0,1)} \e^{\i t\epsilon_e^{(s)}(\lambda)} {\mathcal Q}_e^{(s)} +\bbbone -\sum_{(e,s)\neq (0,1)} {\mathcal Q}_e^{(s)}\nonumber\\
	&=& \sum_{(e,s)\neq (0,1)} \e^{\i t\epsilon_e^{(s)}(\lambda)} {\mathcal Q}_e^{(s)} +{\rm tr} \big(\rho_{\s,\beta,0}\, \cdot\, \big).
	\label{86}
\end{eqnarray}
Combining \eqref{86} with \eqref{83} (with error $\propto\lambda^2$ due to the form of the initial condition) gives 
\begin{equation}
\omega_0\big(\alpha_\lambda^t(X\otimes{\bbbone}_\r)\big ) = {\rm tr}_\s\big( (\rho_{\s,\beta,\lambda}  -\rho_{\s,\beta,0} )X\big) +   \omega_0\big(\e^{\i t {\mathcal M}(\lambda) }(X)\otimes\bbbone_\r \big) +O(\lambda^2 \e^{-\gamma(\lambda)t}). 
	\label{84}
\end{equation}
The first term on the right side is $O(\lambda^2)$, hence
\begin{equation}
	\omega_0\big(\alpha_\lambda^t(X\otimes{\bbbone}_\r)\big ) =  \omega_0\big(\e^{\i t {\mathcal M}(\lambda) }(X)\otimes\bbbone_\r \big) +O(\lambda^2),
	\label{87}
\end{equation}
where the remainder is uniform in $t$. Equation \eqref{87} gives an approximation of the Heisenberg system dynamics by the semigroup $\e^{\i t {\mathcal M}(\lambda)}$, up to a precision $O(\lambda^2)$, for all times. Notice that the state $\omega_{\s,\beta}\otimes\omega_{\r,\beta}$, where $\omega_{\s,\beta}$ is given by the system equilibrium state $\rho_{\s,\beta,0}$, is invariant under this dynamics (see \eqref{85}). We now show that if we truncate the generator ${\mathcal M}(\lambda)$ by taking into account only the part up to $O(\lambda^2)$ in the eigenvalues $\epsilon_e^{(s)}(\lambda)$ in \eqref{67}, then we obtain a CPT semigroup. Using that \footnote{We have $
	\e^{\i t\epsilon(\lambda)} = \e^{\i t(e+ \lambda^2 a +O(\lambda^4))} = \e^{\i t(e+ \lambda^2 a)} + \e^{\i t(e+ \lambda^2 a)}[ \e^{\i t O(\lambda^4)}-1]
	$ and 
	$|\e^{\i t O(\lambda^4)}-1|  = |\i O(\lambda^4) \int_0^t e^{\i s O(\lambda^4)}ds| \le C\lambda^4 t e^{t \lambda^4c}$, for some $C,c>0$ independent of $\lambda, t$. }
\begin{equation}
	\label{19}
	\e^{\i t \epsilon_e^{(s)}(\lambda)} = \e^{\i t(e+ \lambda^2 a_e^{(s)})} + O\big(\lambda^4 t \e^{-\lambda^2 t(\gamma_{{\rm FGR}} +O(\lambda^2))}\big),
\end{equation} 
we obtain
\begin{eqnarray}
\big(\e^{\i t{\mathcal M}(\lambda)}(X)\otimes \bbbone_\s \big)\Omega_{\s,\beta}
&=& \sum_{e,s} \e^{\i t \epsilon_e^{(s)}} Q_e^{(s)}(X\otimes\bbbone_\s) \Omega_{\s,\beta} \nonumber\\
	&=&  \e^{\i t (L_\s+\lambda^2 \Lambda)} (X\otimes\bbbone_\s)\Omega_{\s,\beta}
+\  O\big(\lambda^4 t \e^{-\lambda^2 t(\gamma_{{\rm FGR}} +O(\lambda^2))}\big),\qquad
		\label{88}
\end{eqnarray}
where the total level shift operator is defined to be 
\begin{equation}
	\label{80}
	\Lambda = \bigoplus_{e\in{\rm spec}(L_\s)}\Lambda_e,
\end{equation}
with $\Lambda_e$ given in \eqref{61}. We now define the group $\delta^t_\lambda$, acting on system observables, by 
\begin{equation}
	\label{81}
	\big(\delta_\lambda^t(X)\otimes\bbbone_\s \big) \Omega_{\s,\beta} = \e^{\i t (L_\s+\lambda^2\Lambda)} (X\otimes\bbbone_\s)\Omega_{\s,\beta}.
\end{equation}
Combining \eqref{87} and \eqref{88} we get, for $\gamma_{\rm FGR}>0$, 
\begin{equation}
	\label{89}
	\omega_0\big(\alpha_\lambda^t(X\otimes{\bbbone}_\r)\big ) =  \omega_0\big(\delta_\lambda^t(X)\otimes\bbbone_\r \big) +O(\lambda^2).
\end{equation}
By duality, we have ${\rm tr}_\s( \rho\, \delta^t_\lambda(X)) = {\rm tr}_\s((e^{t{\mathcal G}}\rho)X)$ for all system density matrices $\rho$ and all system observables $X$. We have $\e^{\i t L_\s}(X\otimes\bbbone_\s)\Omega_{\s,\beta} = (\e^{\i t H_\s} X\e^{-\i t H_\s}\otimes\bbbone_\s)\Omega_{\s,\beta}$, which follows simply from $\e^{-\i t L_\s}\Omega_{\s,\beta}=\Omega_{\s,\beta}$. This gives a contribution $-\i[H_\s,\cdot]$ to the generator $\mathcal G$. For nonzero $\lambda$, we then get  ${\mathcal G}\rho = -\i[H_\s,\rho] +\lambda^2 K\rho$, with $K$ satisfying \eqref{-1}, see also Appendix \ref{appendixA}.

Since $(L_\s+\lambda^2\Lambda)\Omega_{\s,\beta}=0$ we have $\delta_\lambda^t(\bbbone_\s)=\bbbone_\s$. It remains to prove that $\delta_\lambda^t$ is completely positive.

\subsubsection{Proof that $\delta_\lambda^t$ is CP}
\label{cp1} It follows from \eqref{89} that 
\begin{equation}
	\label{90}
	\lim_{\lambda\rightarrow 0} \omega_0\big(\alpha_\lambda^{t/\lambda^2} \circ \alpha_0^{-t/\lambda^2} (X\otimes{\bbbone}_\r)\big ) =  \omega_0\big(\sigma^t(X)\otimes\bbbone_\r \big),
\end{equation}
where $\sigma^t$ is defined by 
\begin{equation}
	\label{91}
	\big(\sigma^t(X)\otimes\bbbone_\s \big) \Omega_{\s,\beta} = \e^{\i t \Lambda} (X\otimes\bbbone_\s)\Omega_{\s,\beta}.
\end{equation}
Since limits of CP maps are CP, we know from \eqref{90} that $\sigma^t$ is CP. Next, $\delta_\lambda^t$ is the composition of two CP maps,
$$
\delta_\lambda^t = \big( \e^{\i t H_\s} \cdot \e^{-\i t H_\s} \big) \circ \sigma^{\lambda^2 t},
$$
and hence it is CP itself. This shows \eqref{24}.

\subsection{Proof of \eqref{28}}

\subsubsection{The renormalized quantities}

The reduced system equilibrium density matrix $\rho_{\s,\beta,\lambda}$ is defined by the relation
\begin{equation}
	\label{114}
	{\rm tr} \big( \rho_{\s,\beta,\lambda} X\big) =\omega_{\s\r,\beta,\lambda} (X\otimes\bbbone_\r),\qquad \forall X
\end{equation}
where $\omega_{\s\r,\beta,\lambda}$ is the coupled system-reservoir equilibrium state whose purification is \eqref{41}. We introduce the renormalized system Hamiltonian $\widetilde H_\s(\lambda)$ by the relation \eqref{27}. This defines $\widetilde H_\s(\lambda)$ only up to an additive term $\propto \bbbone_\s$. Of course, we would like the property $\widetilde H_\s(0)=H_\s$, which will determine this additive term. Without loss of generality, we suppose that $\min{\rm spec} H_\s=0$ (the smallest eigenvalue of $H_\s$ is normalized to be at the origin). Let $\widetilde E_0(\lambda)$ be the smallest eigenvalue of $\widetilde H_\s(\lambda)$. We have from \eqref{27} that ${\rm tr} (\e^{-\beta \widetilde H_\s(\lambda)})\ \|\rho_{\s,\beta,\lambda}\|  = \e^{-\beta \widetilde E_0(\lambda)}$, where $\|\rho_{\s,\beta,\lambda}\|$ is the operator norm of the density matrix. Then we impose the normalization $\widetilde E_0(\lambda)=0$, which amounts to ${\rm tr} (\e^{-\beta \widetilde H_\s(\lambda)} )= 1/\|\rho_{\s,\beta,\lambda}\|$ and so we define
\begin{equation}
	\label{115}
	\widetilde H_\s(\lambda) = -\frac1\beta \ln \frac{\rho_{\s,\beta,\lambda}}{\|\rho_{\s,\beta,\lambda}\|}.
\end{equation}
By simple perturbation theory we have $\rho_{\s,\beta,\lambda} =\rho_{\s,\beta,0} +O(\lambda^2)$.\footnote{The correction linear in $\lambda$ vanishes, since in our models, the interaction is linear in the field (c.f. \eqref{1}) and $\scalprod{\Omega_\r}{\varphi_\beta(g)\Omega_\r}=0$.}  It follows from \eqref{115} that 
\begin{equation}
	\label{116}
	\widetilde H_\s(\lambda) = H_\s +O(\lambda^2),
\end{equation}
where $H_\s$ is the original, uncoupled system Hamiltonian \eqref{2}. The spectral representation of the renormalized Hamiltonian is
\begin{equation}
	\label{117}
	\widetilde H_\s(\lambda) = \sum_{j=1}^N \widetilde E_j  |\widetilde \phi_j\rangle\langle\widetilde \phi_j|,
\end{equation}
where $\widetilde E_j$ and $\widetilde \phi_j$ depend on $\lambda$ and satisfy
\begin{equation}
	\label{118}
	|E_j - \widetilde E_j(\lambda)| =O(\lambda^2),\qquad \|\phi_j -\widetilde\phi_j(\lambda)\| = O(\lambda^2).
\end{equation}
In analogy with \eqref{44} we introduce the Liouvillians
\begin{eqnarray}
	\widetilde L_0 &=& \widetilde L_\s + L_\r  \nonumber\\
	\widetilde L_\s &=& \widetilde H_\s\otimes\bbbone_\s - \bbbone_\s\otimes {\mathcal C} \widetilde H_\s\, {\mathcal C} \nonumber\\
	L_\r &=& H_\r\otimes\bbbone_\r -\bbbone_\r\otimes H_\r
	\label{119}
\end{eqnarray}
where ${\mathcal C}$ is the operator taking complex conjugation of coordinates in the basis of eigenvectors $\{\phi_j\}$ of $H_\s$. A purification of $\rho_{\s,\beta,\lambda}$ is given by the vector ($\widetilde Z$ is a normalization constant)
\begin{equation}
	\label{120}
	\widetilde \Omega_{\s,\beta,\lambda} = \widetilde Z^{-1/2} \sum_{j=1}^N \e^{-\beta \widetilde E_j/2 }\widetilde \phi_j\otimes {\mathcal C} \widetilde\phi_j.
\end{equation}
Namely, for any system observable $X$, we have 
\begin{equation}
	\label{123}
	{\rm tr}_\s (\rho_{\s,\beta,\lambda} X) = \langle \widetilde \Omega_{\s,\beta,\lambda}, (X\otimes\bbbone_\s)\widetilde \Omega_{\s,\beta,\lambda}\rangle.
\end{equation} 
We also define
\begin{equation}
	\label{121}
	\widetilde \Omega_0 = \widetilde\Omega_{\s,\beta,\lambda}\otimes\Omega_\r,
\end{equation}
where $\Omega_\r$ is the vacuum \eqref{38}. It is clear from the definitions \eqref{119}, \eqref{120} and \eqref{121} that 
\begin{equation}
	\label{122}
	\widetilde L_\s \widetilde \Omega_{\s,\beta,\lambda} =0\qquad \mbox{and}\qquad \widetilde L_0\widetilde\Omega_0=0.
\end{equation}

Given an eigenvalue $\widetilde e$ of $\widetilde L_0$ (the eigenvalues of $\widetilde L_0$ and of $\widetilde L_\s$ are the same\footnote{\blue{The spectrum of $L_\s$ covers the whole real line and is absolutely continuous except for a simple eigenvalue at the origin (with eigenvector $\Omega_\r$). So the eigenvectors of $\widetilde L_0$ are exactly $\Psi_\s\otimes\Omega_\r$, where $\Psi_\s$ are eigenvectors of $\widetilde L_\s$. And these eigenvectors correspond to the same eigenvalues of the two operators. Of course, the whole spectra do not coincide: $\widetilde L_0$ has additionally continuous spectrum covering $\mathbb R$.
}}), we denote by $\widetilde P_{\widetilde e}$ the associated spectral projection and we define the level shift operators (compare with \eqref{59}, \eqref{80}) 
\begin{equation}
	\label{124}
	\widetilde \Lambda_{\widetilde e} = - \widetilde P_{\widetilde e} I \widetilde P_{\widetilde e}^\perp (\widetilde L_0 -\widetilde e +\i 0)^{-1} I \widetilde P_{\widetilde e}, \ \  \widetilde \Lambda = \bigoplus_{\widetilde e\in{\rm spec}(\widetilde L_\s)}\Lambda_{\widetilde e}.
\end{equation}
A perturbation theory argument based on \eqref{116} shows that $\widetilde \Lambda_{\widetilde e}-\Lambda_e =O(\lambda^2)$.  Assuming that the $\Lambda_e$ have the expansion \eqref{61} (where all $a_e^{(s)}$ are distinct, for simplicity), the operator $\widetilde \Lambda_{\widetilde e}$ has a similar expansion,
\begin{equation}
	\widetilde \Lambda_{\widetilde e} = \sum_{s=1}^{m_e} \widetilde a_{\widetilde e}^{(s)} \widetilde Q_{\widetilde e}^{(s)},
\end{equation} 
where $\widetilde a_{\widetilde e}^{(s)}$ and $\widetilde Q_{\widetilde e}^{(s)}$ are the eigenvalues and rank-one eigenprojections, satisfying
\begin{equation}
	\label{126}
	a_e^{(s)} = \widetilde a_{\widetilde e}^{(s)} +O(\lambda^2),\qquad \widetilde Q_{\widetilde e}^{(s)} =Q_e^{(s)} +O(\lambda^2).
\end{equation}
One also shows that (compare with \eqref{125}, and see \cite{KoMeCP}, Proposition 3.2)
\begin{equation}
	\widetilde \Lambda_0 \widetilde \Omega_0 =0,\quad \mbox{i.e.,}\quad  \widetilde a_0^{(1)} =0,\quad \widetilde Q_0^{(1)} = |\widetilde \Omega_0\rangle\langle\widetilde \Omega_0|.
	\label{127}
\end{equation}

\subsubsection{The resonance expansion} 

The vector $\widetilde\Omega_0$ is cyclic and separating and furthermore, one can find an operator $D'$, which commutes with all system-reservoir observables \footnote{Some care has to be taken here as $D'$ is not a bounded operator, but the technicalities of this difficulty are not too severe to overcome, see Lemma 3.4 of \cite{KoMeCP}.}, and which satisfies 
\begin{equation}
	\label{103}
	\widetilde\Omega_0 = D'\Omega_{\s\r,\beta,\lambda}, \qquad D'=\bbbone +O(\lambda).
\end{equation}
(The existence of a {\em bounded} $D'$ belonging to the commutant of the operator algebra, and which satisfies \eqref{103} to arbitrary precision, is guaranteed by the separating property of $\Omega_{\s\r,\beta,\lambda}$. However, \eqref{103} is an {\em equality}, not an approximation. The equality can be obtained due to the special form of the vectors involved, see \cite{KoMeCP}.)  We take initial conditions of the form
\begin{equation}
	\label{101}
	\Psi_0 = B' \widetilde\Omega_0 = B'D' \Omega_{\s\r,\beta,\lambda},
\end{equation}
where $B'$ belongs to the commutant (as before) and where the second equality follows from \eqref{103}. Varying over $B'$, the vectors $\Psi_0$ form a dense set. We repeat the argument in \eqref{33},
\begin{eqnarray}
\lefteqn{\!\!\!\!\!\!\!\!\!\!\!\!	\omega_0\big(\alpha^t_\lambda (X\otimes\bbbone_\r)\big) = \scalprod{\Psi_0}{\e^{\i t L_\lambda} (X\otimes\bbbone_\s\otimes\bbbone_\r)\e^{-\i t L_\lambda}\Psi_0}} \nonumber\\
	&=& \scalprod{\Psi_0}{B' D' \e^{\i t L_\lambda}  (X\otimes\bbbone_\s\otimes\bbbone_\r) \Omega_{\s\r,\beta,\lambda}}.
	\label{102}
\end{eqnarray}
Then we perform again the spectral deformation, \eqref{50} and deform the contour of integration, to arrive at (compare with \eqref{54})
\begin{eqnarray}
\omega_0\big(\alpha_\lambda^t(X\otimes\bbbone_\r)\big ) &=& \sum_{e\in{\rm spec}(L_\s)} \sum_{s=1}^{m_e} \e^{\i t \epsilon_e^{(s)}}  \scalprod{[(D'B')^*\Psi_0]_{\bar\theta}}{\Pi_e^{(s)} \big( X\otimes\bbbone_\s\otimes \bbbone_\r \big) [\Omega_{\s\r,\beta,\lambda}]_\theta}\nonumber\\
	&&+ \, O\big( \lambda \e^{- \frac 34\theta_0 t}\big).
		\label{104}
\end{eqnarray}
The term $e=0$, $s=1$ is (see \eqref{53})
\begin{eqnarray}
	\lefteqn{
\scalprod{[(D'B')^*\Psi_0]_{\bar\theta}}{[\Omega_{\s\r,\beta,\lambda}]_\theta}  \scalprod{[\Omega_{\s\r,\beta,\lambda}]_{\bar\theta} }{\big( X\otimes \bbbone_\s\otimes \bbbone_\r \big) [\Omega_{\s\r,\beta,\lambda}]_\theta} } \nonumber\\
	&\qquad\qquad=& {\rm tr}_\s \big( \rho_{\s,\beta,\lambda} X\big)\nonumber\\
	& \qquad\qquad=& \scalprod{\widetilde \Omega_0}{\big(X\otimes\bbbone_\s \otimes \bbbone_\r\big)\widetilde\Omega_0}\nonumber\\ &\qquad\qquad=&\scalprod{(B')^*\Psi_0}{ |\widetilde\Omega_0\rangle \langle\widetilde\Omega_0| \big(X\otimes\bbbone_\s \otimes\bbbone_\r\big)\widetilde \Omega_0}.
	\label{105}
\end{eqnarray}
We use here that $\scalprod{[(D'B')^*\Psi_0]_{\bar\theta}}{[\Omega_{\s\r,\beta,\lambda}]_\theta} =1$ and $\langle (B')^*\Psi_0, \widetilde\Omega_0\rangle =1$. In the other terms, $(e,s)\neq (0,1)$, in the sum in \eqref{104}, we replace $D'$ by $\bbbone$ (see \eqref{103}), use the approximation \eqref{63} and retain only the part $e+\lambda^2 a_e^{(s)}$ in the resonance energies (see \eqref{19}). Then \eqref{104} and \eqref{105} give  (see also \eqref{19})
\begin{eqnarray}
	\omega_0\big(\alpha_\lambda^t(X\otimes\bbbone_\r)\big )
 &=& \scalprod{(B')^*\Psi_0}{ |\widetilde\Omega_0\rangle \langle\widetilde\Omega_0| \big(X\otimes\bbbone_\s \otimes\bbbone_\r\big)\widetilde \Omega_0} \nonumber\\
	&& + \sum_{(e,s)\neq (0,1)} \e^{\i t (e +\lambda^2 a_e^{(s)}) }  \scalprod{(B')^*\Psi_0}{Q_e^{(s)} \big( X\otimes \bbbone_\s\otimes \bbbone_\r \big) \Omega_{\s\r,\beta,\lambda}}\nonumber\\
	&& + O\Big( (\lambda+\lambda^4t)  \e^{- \lambda^2 t (\gamma_{\rm FGR} + O(\lambda^2))}\Big)\nonumber\\
	&&+O\big( \lambda \e^{- \frac 34\theta_0 t}\big).
	\label{106}
\end{eqnarray}
Next, since $e+ \lambda^2 a_e^{(s)} = \widetilde e +\lambda^2 \widetilde a_{\widetilde e}^{(s)} +O(\lambda^2)$  and $Q_e^{(s)} = \widetilde Q_{\widetilde e}^{(s)} +O(\lambda^2)$ (see \eqref{126}), we replace in \eqref{106} $e+ \lambda^2 a_e^{(s)}$ and $Q_e^{(s)}$ by $\widetilde e +\lambda^2 \widetilde a_{\widetilde e}^{(s)}$ and $\widetilde Q_{\widetilde e}^{(s)}$,  incurring an error $O((\lambda + \lambda^2 t)\e^{-\lambda^2 t\gamma_{\rm FGR}})$ (proceed similarly as in \eqref{19}). But now,
\begin{equation}
	\label{107}
	\sum_{(\widetilde e,s)\neq (0,1)} \e^{\i t (\widetilde e +\lambda^2\widetilde a_{\widetilde e}^{(s)})} \widetilde Q_{\widetilde e}^{(s)} = \e^{\i t (\widetilde L_\s+ \lambda^2 \widetilde \Lambda)} P(\widetilde \Lambda \neq 0)
\end{equation}
and $P(\widetilde\Lambda=0) = |\widetilde \Omega_0\rangle\langle \widetilde \Omega_0|$, 
where $P(\widetilde\Lambda\neq 0)$ and $P(\widetilde\Lambda= 0)$ are spectral (Riesz) projections. 
(See also \eqref{124} and \eqref{127}.) Therefore, the two main terms on the right side of \eqref{106} yield the operator $\e^{\i t(\widetilde L_\s+\lambda^2 \widetilde\Lambda)}$, namely, 
\begin{eqnarray}
\lefteqn{	\omega_0\big(\alpha_\lambda^t(X\otimes \bbbone_\r)\big )}\nonumber\\
 &=& \scalprod{(B')^*\Psi_0}{ \e^{\i t (\widetilde L_\s+\lambda^2 \widetilde \Lambda)} \big(X\otimes\bbbone_\s \otimes\bbbone_\r\big)\widetilde \Omega_0} \nonumber\\
	&&	+O\Big( (\lambda+\lambda^2t)  \e^{- \lambda^2 t (\gamma_{\rm FGR} +O(\lambda^2))}\Big).
	\label{108}
\end{eqnarray}
By cyclicity of $\widetilde\Omega_{\s,\beta,\lambda}$, the relation
\begin{equation}
	\label{109}
	\e^{\i t (\widetilde L_\s+\lambda^2 \widetilde \Lambda)} \big(X\otimes\bbbone_\s \big)\widetilde \Omega_{\s,\beta,\lambda} = \big( \tau_\lambda^t(X)\otimes \bbbone_\s\big) \widetilde \Omega_{\s,\beta,\lambda}
\end{equation}
defines uniquely a group (in $t$), $\tau_\lambda^t$,  acting on system observables. Using \eqref{109} and commuting $B'$ through the observable and using $B'\widetilde\Omega_0=\Psi_0$, we obtain for the first term on the right side of \eqref{108} simply the expression $\langle\Psi_0, (\tau^t_\lambda(X)\otimes\bbbone_\s\otimes\bbbone_\r)\Psi_0\rangle = \omega_0( \tau^t_\lambda(X)\otimes\bbbone_\r )$. So  \eqref{108} yields
\begin{equation}
\omega_0\big(\alpha_\lambda^t(X\otimes \bbbone_\r)\big )  = \omega_0\big( \tau^t_\lambda(X)\otimes\bbbone_\r \big)
+O\Big( (\lambda+\lambda^2t)  \e^{- \lambda^2 t (\gamma_{\rm FGR} +O(\lambda^2))}\Big).
		\label{111}
\end{equation}
For initial states $\omega_0=\omega_\s\otimes\omega_{\r,\beta}$, where $\omega_\s$ is given by a density matrix $\rho$ and $\omega_{\r,\beta}$ is the reservoir equilibrium (or a local perturbation thereof), we get
\begin{eqnarray}
\omega_0\big(\alpha_\lambda^t(X\otimes \bbbone_\r)\big )  = {\rm tr}_\s\big(\rho  \tau^t_\lambda(X)\big)
 +O\Big( (\lambda+\lambda^2t)  \e^{- \lambda^2 t (\gamma_{\rm FGR} +O(\lambda^2))}\Big).
		\label{112}
\end{eqnarray}
By duality, we define uniquely $M(\lambda)$, an operator acting on system density matrices, by
\begin{equation}
	\label{113}
	{\rm tr}_\s\big(\rho \tau^t_\lambda(X)\big) = {\rm tr}\big( (\e^{ t M(\lambda)}\rho) X\big),
\end{equation}
and \eqref{28} follows from \eqref{112}, \eqref{113}. 

That $\tau^t_\lambda(\bbbone_\s)=\bbbone_\s$ is clear from the definition \eqref{109}, as $(\widetilde L_\s+\lambda^2\widetilde \Lambda)\widetilde \Omega_{\s,\beta,\lambda}=0$. We show below in Section \ref{subsubcp}  that for  $\lambda, t$ fixed,  $\tau^t_\lambda$ is a CP map.

\smallskip

{\em Evolution of observables $X$ commuting with $H_\s$. \ }
We treat the general term in the sum of \eqref{106} as follows,
\begin{eqnarray}
	\e^{\i t e} Q_e^{(s)} (X\otimes\bbbone_\s\otimes\bbbone_\r) \Omega_{\s\r,\beta,\lambda}
 &=& Q_e^{(s)} \e^{\i t L_\s} (X\otimes\bbbone_\s\otimes\bbbone_\r) \Omega_{\s\r,\beta,\lambda} \nonumber\\
	&=& Q_e^{(s)} \e^{\i t L_\s} (X\otimes\bbbone_\s\otimes\bbbone_\r) \Omega_{\s\r,\beta,0} + O(\lambda) \nonumber\\
	&=& Q_e^{(s)}  (X_t\otimes\bbbone_\s\otimes\bbbone_\r) \Omega_{\s\r,\beta,0} +O(\lambda) \nonumber\\
	&=& Q_e^{(s)}  (X_t\otimes\bbbone_\s\otimes\bbbone_\r) \Omega_{\s\r,\beta,\lambda} +O(\lambda).
	\label{mh1}
\end{eqnarray}
Here, we have set
\begin{equation}
	X_t \equiv \e^{\i t H_\s} X \e^{-\i t H_\s}.
	\label{mh2}
\end{equation}
The first equality in \eqref{mh1} is due to \eqref{62}. The third one comes from $\e^{-\i t L_\s}\Omega_{\s\r,\beta,0}=\Omega_{\s\r,\beta,0}$ and the remaining ones follow from $\Omega_{\s\r,\beta,\lambda} - \Omega_{\s\r,\beta,0} = O(\lambda)$. We now use \eqref{mh1} in the sum over $(e,s)\neq (0,1)$ in \eqref{106} and arrive at 
\begin{eqnarray}
	\omega_0\big(\alpha_\lambda^t(X\otimes \bbbone_\r)\big )
 &=& \scalprod{(B')^*\Psi_0}{ |\widetilde\Omega_0\rangle \langle\widetilde\Omega_0| \big(X\otimes\bbbone_\s \otimes\bbbone_\r\big)\widetilde \Omega_0}\nonumber\\
 &&  + \sum_{(e,s)\neq (0,1)} \e^{\i t \lambda^2 a_e^{(s)} } \scalprod{(B')^*\Psi_0}{Q_e^{(s)} \big( X_t\otimes{\bbbone}_\s\otimes {\bbbone}_\r \big) \Omega_{\s\r,\beta,\lambda}}\nonumber\\
	&& + O\Big( (\lambda+\lambda^4t)  \e^{- \lambda^2 t (\gamma_{\rm FGR} + O(\lambda^2))}\Big)+O\big( \lambda \e^{- \frac 34\theta_0 t}\big).
	\label{106.1}
\end{eqnarray}
Replacing in the last sum $\e^{\i t\lambda^2 a_e^{(s)}}$ by $\e^{\i t\lambda^2 \widetilde a_{\widetilde e}^{(s)}}$ we  incur an error of $O(\lambda^4t \e^{-\lambda^2t(\gamma_{\rm FGR}+O(\lambda^2))})$. Now we define the group $\tau_{{\rm d},\lambda}^t$,  acting on system observables, by
\begin{equation}
	\label{109.1}
	\e^{\i t \lambda^2 \widetilde \Lambda} \big(X\otimes\bbbone_\s \big)\widetilde \Omega_{\s,\beta,\lambda} = \big( \tau_{{\rm d},\lambda}^t(X)\otimes \bbbone_\s\big) \widetilde \Omega_{\s,\beta,\lambda}.
\end{equation}
Combining \eqref{109.1} with \eqref{106.1} then yields \blue{ (recall also \eqref{105})
\begin{eqnarray}
	\omega_0\big(\alpha_\lambda^t(X\otimes \bbbone_\r)\big )
	&=&  {\rm tr}_\s\Big(\rho_{\s,\beta,\lambda}  \big(X-X_t\big)\Big) +  \omega_0\big(\tau_{{\rm d},\lambda}^t(X_t)\otimes\bbbone_\r \big)\nonumber\\
	&& + O\Big( (\lambda+\lambda^4t)  \e^{- \lambda^2 t (\gamma_{\rm FGR} + O(\lambda^2))}\Big).
	\label{106.3}
\end{eqnarray}
Assuming that the initial state $\omega_0$ of product form we can express \eqref{106.3} in the dual space as relation \eqref{106.4}. For the invariant observables $X$ s.t. $[X,H_\s]=0$, we have $X_t=X$ for all $t$, so
}
\begin{equation}
\omega_0\big(\alpha_\lambda^t(X\otimes \bbbone_\r)\big ) = \omega_0\big(\tau_{{\rm d},\lambda}^t(X)\otimes\bbbone_\r \big)+ O\Big( (\lambda+\lambda^4t)  \e^{- \lambda^2 t (\gamma_{\rm FGR} + O(\lambda^2))}\Big).
	\label{106.2}
\end{equation}
It is clear from \eqref{109.1} and \eqref{127} that $\tau_{{\rm d},\lambda}^t(\bbbone_\s)=\bbbone_\s$. We show below in Section \ref{subsubcp} that $\tau_{{\rm d},\lambda}^t$ is completely positive. 
Again by duality, and for an initial condition $\omega_0={\rm tr}_\s (\rho\,  \cdot)\otimes\omega_{\r,\beta}$, equation \eqref{106.2} becomes
\begin{equation}
{\rm tr}_\s(V_t\rho) X = {\rm tr}_\s (\e^{t\lambda^2 M_{\rm d}(\lambda)}\rho) X+ O\Big( (\lambda+\lambda^4t)  \e^{- \lambda^2 t (\gamma_{\rm FGR} + O(\lambda^2))}\Big),
	\label{mh3}
\end{equation}
valid $\forall X {\rm \ s.t.\ } [X,H_\s]=0$.
Taking $X=|\varphi_k\rangle\langle\varphi_k|$ we obtain equation \eqref{mh4}.

\subsubsection{Proof that $\tau^t_\lambda$ and $\tau^t_{{\rm d},\lambda}$ are CP} 
\label{subsubcp} The idea is to view $\tau^t_\lambda$ as a weak coupling dynamics and proceed as in Subsection \ref{cp1}. To do this, introduce the Liouvillian
\begin{equation}
	\widetilde L_\mu = \widetilde L_0  + \mu\lambda I,
	\label{130}
\end{equation}
where $\widetilde L_0$ is given in \eqref{119} and the interaction $I$ is \eqref{45}. Here we consider $\mu\in\mathbb R$ as the interaction constant, and $\lambda$ is viewed as part of the interaction operator. (Recall that $\widetilde L_0$ also depends on $\lambda$.) The eigenvalues of the unperturbed $\widetilde L_\mu |_{\mu=0}$ are the same as those of $\widetilde L_0$ and the levels shift operators associated to \eqref{130} are given by \eqref{124} with $I$ replaced by $\lambda I$ (they give the quadratic corrections in $\mu$  to the spectrum). In other words, $\lambda^2 \widetilde \Lambda$, with $\widetilde \Lambda$ given in \eqref{124}, is the (complete) level shift operator of $\widetilde L_\mu$. We define the dynamics $\widetilde \gamma_\mu^t$ by
\begin{equation}
	\label{129}
	\omega_0\big( \widetilde \gamma^t_\mu (X\otimes{\mathcal P} ) \big) = \scalprod{\Psi_0}{ \e^{\i t \widetilde L_\mu}( X\otimes\bbbone_\s\otimes{\mathcal P}_\beta  ) \e^{-\i t \widetilde L_\mu} \Psi_0}.
\end{equation}
In \eqref{129}, $X$ and ${\mathcal P}$ are system and reservoir observables, with ${\mathcal P}_\beta$ being the representation in the purification space, see also  \eqref{37}.  The equilibrium (KMS) state associated to $\widetilde L_\mu$ is given by (compare with \eqref{41})
\begin{equation}
	\label{132}
	\widetilde \Omega_{\s\r,\beta,\mu}  =\frac{\e^{-\frac{\beta}{2} (\widetilde L_0+\mu \lambda G\otimes\bbbone_\s\otimes\varphi_\beta(g) )} \Omega_{\s\r,\beta,0} }{\|\e^{-\frac{\beta}{2} (\widetilde L_0+\mu \lambda G\otimes\bbbone_\s\otimes\varphi_\beta(g) )} \Omega_{\s\r,\beta,0} \| } 
\end{equation}
(and depends on $\lambda$ as well). This is a cyclic and separating vector and the initial condition can be written as $\Psi_0 = B'D'\widetilde \Omega_{\s\r,\beta,\mu}$ (c.f. \eqref{101}). We then obtain (c.f. \eqref{102})
\begin{equation}
	\label{133}
	\omega_0\big(\widetilde\gamma^t_\lambda (X\otimes\bbbone_\r)\big) 
	= \scalprod{\Psi_0}{B' D' \e^{\i t \widetilde L_\mu}  (X\otimes\bbbone_\s\otimes\bbbone_\r) \widetilde \Omega_{\s\r,\beta,\mu}}
\end{equation}
(with $B'$, $D'$ depending on both $\lambda$ and $\mu$). Proceeding to perform the spectral deformation and resonance expansion in the same manner as we did in Sections \ref{secrepdyn} --\ref{secother}, we obtain (analogous to \eqref{87}),
\begin{equation}
\omega_0\big(\widetilde\gamma_\mu^{t/\mu^2}(X\otimes{\bbbone}_\r)\big ) = \scalprod{(B')^*\Psi_0}{ \e^{\i t (\widetilde L_0+ \mu^2\lambda^2 \widetilde \Lambda)} \big(X\otimes\bbbone_\s \otimes\bbbone_\r\big)\widetilde \Omega_0} +O(\mu^2),
		\label{134}
\end{equation}
with a remainder term uniform in $t$. It follows that 
\begin{eqnarray}
	\lim_{\mu\rightarrow 0}	\omega_0\big(\widetilde\gamma_\mu^{t/\mu^2}\circ\widetilde\gamma_0^{-t/\mu^2} (X\otimes{\bbbone}_\r)\big )
 &=& \scalprod{(B')^*\Psi_0}{ \e^{\i t \lambda^2 \widetilde \Lambda} \big(X\otimes\bbbone_\s \otimes\bbbone_\r\big)\widetilde \Omega_0}\nonumber\\
 &=&\omega_0\big((\tau_{{\rm d},\lambda}^t(X)\otimes\bbbone_\r)\big).
		\label{134.1}
\end{eqnarray}
Consequently, $\tau^t_{{\rm d},\lambda}$ is CP. Since $\tau^t_\lambda = \tau^t_{{\rm d},\lambda}\circ (\e^{\i t \widetilde H_\s} \cdot \e^{-\i t \widetilde H_\s})$ it follows that $\tau^t_\lambda$ is CP as well. 

\blue{
\section{Conclusion}

We establish rigorous bounds on Markovian approximations to the dynamics of a finite dimensional quantum system linearly coupled to an environment of free quantum particles (a quantum field). We show that the Markovian master equation is valid for all times, approximating the true dynamics to $O(\lambda^2)$, $\lambda$ being the system-environment coupling constant. Further, we construct a new Markovian semigroup which is asymptotically exact, meaning that it approximates the true dynamics and converges to the correct final state to all orders in $\lambda$, as time tends to infinity. Our method is based on the quantum dynamical resonance theory which we explain in some detail. In particular, we derive the theory for a wide class of initial system-reservoir states, including entangled states. 
Our approach is purely analytical and our constructions are based on concrete perturbation theory in $\lambda$, valid for all times.
}
\bigskip

{\bf Acknowledgements.\ } The author thanks Martin K\"onenberg for his collaborations on which the present work builds, and for carefully proofreading the current manuscript, and Roberto Floreanini and Philipp Strasberg for valuable comments. \blue{Sincere thanks go to two referees who lent their expertise to examine this work in depth,  and suggested very beneficial clarifications.} The author was supported by the Simons
Foundation and the Centre de Recherches Math\'ematiques, through the Simons-CRM
scholar-in-residence program, as well as by an NSERC Discovery Grant and an NSERC Discovery Accelerator Supplement grant.

\appendix
\section{Explicit form of the generator $K$}
\label{appendixA}
We define the generator ${\mathcal G}$ acting on system density matrices  by
\begin{equation}
	{\rm tr}_\s \big( \rho\,  \delta^t_\lambda(X) \big) = {\rm tr}_\s\big( (e^{t{\mathcal G}}\rho)\,  X\big),
\end{equation}
valid for all system observables and density matrices $X$ and $\rho$. Here, $\delta^t_\lambda$ is given in \eqref{81}. We show that 
\begin{equation}
	{\mathcal G}\rho  = -\i\,  [H_\s, \rho] + \lambda^2 K\rho,
	\label{a2.0}
\end{equation}
where $[\cdot, \cdot]$ is the commutator and, denoting by $\{\cdot, \cdot\}$ the anti-commutator,
\begin{eqnarray}
	K\rho &=& \widehat h(0)\sum_{k,\ell =1}^N \Big(P_k GP_k  \rho P_\ell G P_\ell -\tfrac12 \big\{ P_\ell G P_\ell \, P_k G P_k, \rho \big\}\Big) \nonumber\\
	& & + \sum_{k,\ell\, :\,  k\neq \ell}\widehat h(E_k-E_\ell) \Big( P_\ell GP_k \rho  P_k G P_\ell -\tfrac12 \big\{ P_k G P_\ell G P_k, \rho \big\}\Big)\nonumber\\
	&& - \ \i \, [H_{\rm LS}, \rho ]\label{a3.0}
\end{eqnarray}
and 
\begin{equation}
H_{\rm LS} = \frac1\pi \sum_{k,\ell=1}^N \Big( {\rm P.V.} \int_{{\mathbb R}}  \frac{\widehat h(u)}{E_k-E_\ell -u}\ du\Big) P_k G P_\ell G P_k.\qquad  
	\label{a4.1}
\end{equation}
Here, $\widehat h(u)$ is the Fourier transform of the correlation function,
\begin{equation}
	\widehat h(u) = \int_{{\mathbb R}} e^{- \i tu} \, \omega_{\r,\beta} \big(\varphi(g) \varphi(g_t)\big) dt,\qquad u\in{\mathbb R}
	\label{a5.1}
\end{equation}
where $g(k)$ is the form factor and $g_t(k) = e^{\i \omega(k) t} g(k)$. We have the expression ($u\in\mathbb R$, $\omega\ge 0$)
$$
\widehat h(u) = J_{\rm noise}(|u|) \left| \frac{\e^{\beta u}}{\e^{\beta u}-1}\right|, \ J_{\rm noise}(\omega) = \frac{\pi}{2} \omega^2  \int_{S^2} |g(\omega,\Sigma)|^2 d\Sigma
$$
(spherical coordinates).  $J_{\rm noise}$ is called the reservoir spectral density and $\widehat h(0)$ is understood as the limit $u\rightarrow 0$ of $\widehat h(u)$, \eqref{a5.1}. The first two terms in \eqref{a3.0} constitute the dissipator and the commutator is with the Lamb shift Hamiltonian $H_{\rm LS}$, representing a correction to the system energies. $K$ is the usual Davies generator \cite{AL,BP,CP}. It is manifestly CPT due to the results \cite{GKS,L}.

\medskip
In order to show \eqref{a2.0}-\eqref{a4.1} we first calculate ${\mathcal G}_*$, defined by $\e^{t{\mathcal G}_*}X = \delta^t_\lambda(X)$, {\em i.e.}, 
\begin{equation}
	\big( ({\mathcal G}_*X)\otimes{\bbbone}_\s\big)\Omega_{\s,\beta} = \i (L_\s+\lambda^2 \Lambda) (X\otimes{\bbbone}_\s)\Omega_{\s,\beta}.
	\label{a2}
\end{equation}
The definitions of $L_\s$ and $\Lambda$ are \eqref{44} and \eqref{80}, \eqref{59} and the system Gibbs state $\Omega_{\s,\beta}$ is defined in \eqref{39}. 
For any system operators $X$, $Y$ and $Z$  we have 
\begin{equation}
(Y\otimes \bbbone_\s) \ J_\s (Z\otimes\bbbone_\s)J_\s\, (X\otimes \bbbone_\s)\Omega_{\s,\beta} = \big( (YXe^{-\beta H_\s/2} Z^* e^{\beta H_\s/2})\otimes\bbbone_\s\big)\Omega_{\s,\beta},
	\label{a3}
\end{equation}
\blue{where $J_\s$ is the system modular conjugation \eqref{m6}. To verify \eqref{a3} we first note that by \eqref{m4} we have $J_\s L_\s J_\s = -L_\s$ and so $J_\s=J_\s \e^{\beta L_\s/2} \e^{-\beta L_\s/2} = \e^{-\beta L_\s/2} J_\s\e^{-\beta L_\s/2}$. This together with and \eqref{m3} gives
\begin{equation}
	\label{r5}
J_\s(X\otimes\bbbone_\s)\Omega_{\s,\beta}= \e^{-\beta L_\s/2}  (X^*\otimes\bbbone_\s)\Omega_{\s,\beta} = (\e^{-\beta H_\s/2} X^*\e^{\beta H_\s/2} \otimes \bbbone_\s)\Omega_{\s,\beta}.
\end{equation}
In the last step, we have used that $L_\s=H_\s\otimes\bbbone_\s - \bbbone_\s\otimes H_\s$ and $L_\s\Omega_{\s,\beta}=0$. We now apply \eqref{r5} again to find out the action of $J_\s(Z\otimes\bbbone_\s)$ on the left side of \eqref{a3} and we  easily arrive at the equality \eqref{a3}.} 

It is then clear that $\i L_\s (X\otimes{\bbbone}_\s)\Omega_{\s,\beta} = (\i [H_\s, X]\otimes\bbbone_\r)\Omega_{\s,\beta}$. This gives a contribution $-\i[H_\s,\cdot\, ]$ to $\mathcal G$. To calculate the contribution coming from $\i\lambda^2\Lambda$, we consider the situation where all nonzero eigenvalue differences $e=E_k-E_\ell$ are simple (the general case is done in the same way). Then the projections in \eqref{59} are rank one for $e\neq 0$, $P_e=P_k\otimes P_\ell\otimes |\Omega_\r\rangle \langle \Omega_\r|$, where $P_k=|\phi_k\rangle\langle\phi_k|$ (see \eqref{2}). The projection onto the eigenvalue $e=0$ of $L_\s$ has dimension $N$, $P_{e=0}=\sum_{j=1}^N P_j\otimes P_j\otimes |\Omega_\r\rangle \langle \Omega_\r|$. By expanding $\Lambda_e$, \eqref{59}, using the form \eqref{45} of the interaction $I$ we arrive at the expressions \eqref{a3.0}, \eqref{a4.1}.

\end{document}